\pdfoutput=1
\documentclass[american,preprint,preprintnumbers,prb,showpacs,superscriptaddress]{revtex4-1}

\usepackage{amsmath}
\usepackage{graphicx}
\usepackage{amssymb}
\usepackage{times}
\usepackage{color}
\definecolor{lr}{rgb}{1.0,0.3,0.3}
\definecolor{dg}{rgb}{0.0,0.5,0.0}

\graphicspath{./}
 


\begin{document}

\title{Theoretical model of the dynamic spin polarization of nuclei coupled to paramagnetic point defects in diamond and silicon carbide}

\author{Viktor Iv\'ady}
\affiliation{Department of Physics, Chemistry and Biology, Link\"oping
  University, SE-581 83 Link\"oping, Sweden}
\affiliation{Wigner Research Centre for Physics, Hungarian Academy of Sciences,
  PO Box 49, H-1525, Budapest, Hungary}
  
\author{Kriszti\'an Sz\'asz} 
\affiliation{Wigner Research Centre for Physics, Hungarian Academy of Sciences,
  PO Box 49, H-1525, Budapest, Hungary}

\author{Abram L. Falk}
\affiliation{Institute for Molecular Engineering, University of Chicago, Chicago, IL , USA}
\affiliation{IBM T.J. Watson Research Center, Yorktown Heights, NY, USA}

\author{Paul V. Klimov}
\affiliation{Institute for Molecular Engineering, University of Chicago, Chicago, IL , USA}
\affiliation{Department of Physics, University of California, Santa Barbara, CA, USA}

\author{David J. Christle}
\affiliation{Institute for Molecular Engineering, University of Chicago, Chicago, IL , USA}
\affiliation{Department of Physics, University of California, Santa Barbara, CA, USA}

\author{Erik Janz\'en}
\affiliation{Department of Physics, Chemistry and Biology, Link\"oping
  University, SE-581 83 Link\"oping, Sweden}

\author{Igor A. Abrikosov}
\affiliation{Department of Physics, Chemistry and Biology, Link\"oping
  University, SE-581 83 Link\"oping, Sweden}
\affiliation{Materials Modeling and Development Laboratory, National University of Science and Technology `MISIS', 119049 Moscow, Russia}
\affiliation{LACOMAS Laboratory, Tomsk State University, 634050 Tomsk, Russia}

\author{David D. Awschalom}
\affiliation{Institute for Molecular Engineering, University of Chicago, Chicago, IL , USA}

\author{Adam Gali} 
\email{gali.adam@wigner.mta.hu}
\affiliation{Wigner Research Centre for Physics, Hungarian Academy of Sciences,
  PO Box 49, H-1525, Budapest, Hungary}
\affiliation{Department of Atomic Physics, Budapest University of
  Technology and Economics, Budafoki \'ut 8., H-1111 Budapest,
  Hungary}


\begin{abstract}
Dynamic nuclear spin polarization (DNP) mediated by paramagnetic point defects in semiconductors is a key resource for both initializing nuclear quantum memories and producing nuclear hyperpolarization. DNP is therefore an important process in the field of quantum-information processing, sensitivity-enhanced nuclear magnetic resonance, and nuclear-spin-based spintronics. DNP based on optical pumping of point defects has been demonstrated by using the electron spin of nitrogen-vacancy (NV) center in diamond, and more recently, by using divacancy and related defect spins in hexagonal silicon carbide (SiC). Here, we describe a general model for these optical DNP processes that allows the effects of many microscopic processes to be integrated. Applying this theory, we gain a deeper insight into dynamic nuclear spin polarization and the physics of diamond and SiC defects. Our results are in good agreement with experimental observations and provide a detailed and unified understanding. In particular, our findings show that the defects’ electron spin coherence times and excited state lifetimes are crucial factors in the entire DNP process. 
\end{abstract}
\maketitle

\section{Introduction}

Point defects in solids are promising implementations of quantum bits for quantum computing~\cite{Ladd:Nature2010, Awschalom2013}.  In particular, the negatively charged nitrogen-vacancy defect (NV center) in diamond~\cite{duPreez:1965} has become a leading system in solid-state quantum-information processing because of its unique magnetooptical properties, including long spin coherence times~\cite{Balasubramanian:NatMat2009} and the ease of optical initialization and readout of its spin state,\cite{Jelezko:PSSa2006} even non-destructively~\cite{Awschalom:Nature2010,Robledo:Nature2011}. Since it has a high-spin electronic structure similar to the NV center in diamond, the divacancy in silicon carbide (SiC) has also been proposed to serve as a solid-state quantum bit~\cite{Gali10}. In fact, SiC hosts many other color centers that may also act as quantum bits~\cite{Baranov2005,Baranov2007,Gali10,Weber10,Gali11pss,Gali12jmr}. Recent demonstrations have shown coherent manipulation of divacancy and related defect spins in 4H-\cite{Koehl11}, 6H-\cite{Falk2013,Klimov2014} and 3C-SiC \cite{Falk2013}. Coherent control of the electronic spin of the negatively charged Si vacancy has also been investigated~\cite{Soltamov12,Gali12jmr,NatPhys14}. Further milestones on the path towards robust SiC-based quantum-information technology have been the findings that isolated divacancy qubits have $\sim$1~ms at low temperatures \cite{Christle2014} and that isolated Si-vacancy qubits can operate at room temperature \cite{Widmann2014}.

Coherent control of the electron spins of paramagnetic point defects makes it possible to control and manipulate other spins in the vicinity of the point defect. For instance, the proximate nuclear spins of the NV center in diamond can be polarized~\cite{GSLAC1993,Fuchs2008, Jacques2009,Smeltzer2009,Gali2009,Smeltzer2011,Dreau2012,Fischer2013}, which can be a basis for quantum memories \cite{Dutt:Science2007,Cappellaro:PRL2009,Fuchs2011,Maurer2012}, entanglement-based metrological devices\cite{Giovannetti2011}, and solid-state nuclear gyroscopes \cite{Ajoy2012, Ledbetter2012}. 
A recent demonstration has shown that nuclear spins proximate to divacancies and related defects in 4H- and 6H-SiC can be effectively polarized~\cite{FalkPRL2015}, an important step towards enabling long-lived quantum-information processing in this technologically mature semiconductor material. The transfer of the point defects' electron spin polarization can also lead to hyperpolarization of the host material, thereby enabling sensitivity-enhanced nuclear magnetic resonance and spintronic applications\cite{Terreno2010, Cassidy2013,HaiJing2013,Puttisong2013}

Many of these applications rely on dynamic nuclear spin polarization (DNP) to mediate polarization transfer from the electron spin to neighboring nuclear spins through the hyperfine interaction. Therefore a fundamental understanding of DNP processes is an important aspect of the technological development of nuclear spintronics.

Jacques \emph{et al.}\ \cite{Jacques2009} developed an insightful spin Hamiltonian model that describes the polarization process of $^{15}$N nuclei at the avoided crossing of the diamond NV center’s spin sublevels, otherwise known as the level anti-crossing (LAC). The transverse part of the hyperfine interaction is responsible for the exchange of electron and nuclear polarization \cite{Gali2009}, while the spin selective non-radiative decay of the electron from the excited state (ES) is responsible for the maintenance of the electron-spin polarization. Continuous cycling of optical electronic excitation, flip-flops of electronic and nuclear spins, and non-radiative decay results in a polarization of both the electron and the nuclear spin populations. The so-called ``excited-state level anticrossing (ESLAC) mechanism'' is when spin flip-flops predominantly occur in the excited state. At larger magnetic fields, when the LAC occurs in the ground state (GS), the analogous  ground-state level anticrossing (GSLAC) process occurs \cite{GSLAC1993,HaiJing2013}. To understand the particular features observed in the GS dynamic nuclear spin polarization of a $^{13}$C nuclei adjacent to the vacancy of the NV-center, Wang \emph{et al}.\ \cite{HaiJing2013} recently developed a model capable of describing general hyperfine interactions, i.e.\ for nuclei not on the symmetry axis. Furthermore, to take into account the effect of external strain and spin relaxation and dephasing processes, Fischer \emph{et al}.\ \cite{Fischer2013} recently proposed a density-matrix model.

While these models capture certain phenomena of DNP, a more general model showing predictive power over several color-center systems would be an important development. Here, we propose an extended model that (i) handles general hyperfine interactions of paramagnetic defects with symmetrically or non-symmetrically placed nuclei spins, (ii) takes into account simultaneous GSLAC and ESLAC processes, and (iii) tracks the evolution of spins with time explicitly parameterized. Our model provide insight into the phenomena of the dynamic nuclear spin polarization mechanism for both the NV center in diamond and the divacancies in SiC, and explains several experimental observations. Throughout our investigation, \emph{electron-spin decoherence} appears as an important limiting factor for the polarizability of the nuclear spins. We show that considering electron-spin coherence will be vital to maximizing the performance of DNP in practical applications. 

In Section~\ref{sec:defects}, we briefly describe the electronic structure and the corresponding spin properties of the ground and excited states of the considered point defects, namely, the NV center in diamond and the divacancy defects in 4H- and 6H-SiC. These defects' structures are then used to define the spin Hamiltonian, which is given together with the model of the dynamic nuclear spin polarization in Section~\ref{sec:method}. In Section~\ref{sec:method}, we also summarize the basic concept and parameters of the model to calculate the nuclear spin polarization as a function of different variables. In Section~\ref{sec:abinitio}, we describe the \emph{ab initio} methods and models to calculate the spin related properties of NV center in diamond and divacancy defects in SiC, along with the corresponding results. The full hyperfine tensors of the studied nuclear spins are calculated both in the ground and excited states and are important parameters in the DNP model. In Section~\ref{sec:dnp}, we provide the results and an analysis of DNP for the NV center in diamond and the divacancy in SiC. Finally, we summarize our findings in Section~\ref{sec:conclusion}.

\section{Electronic structure of NV center in diamond and divacancy in SiC}
\label{sec:defects}

\begin{figure}[h!]
\includegraphics[width=0.9\columnwidth]{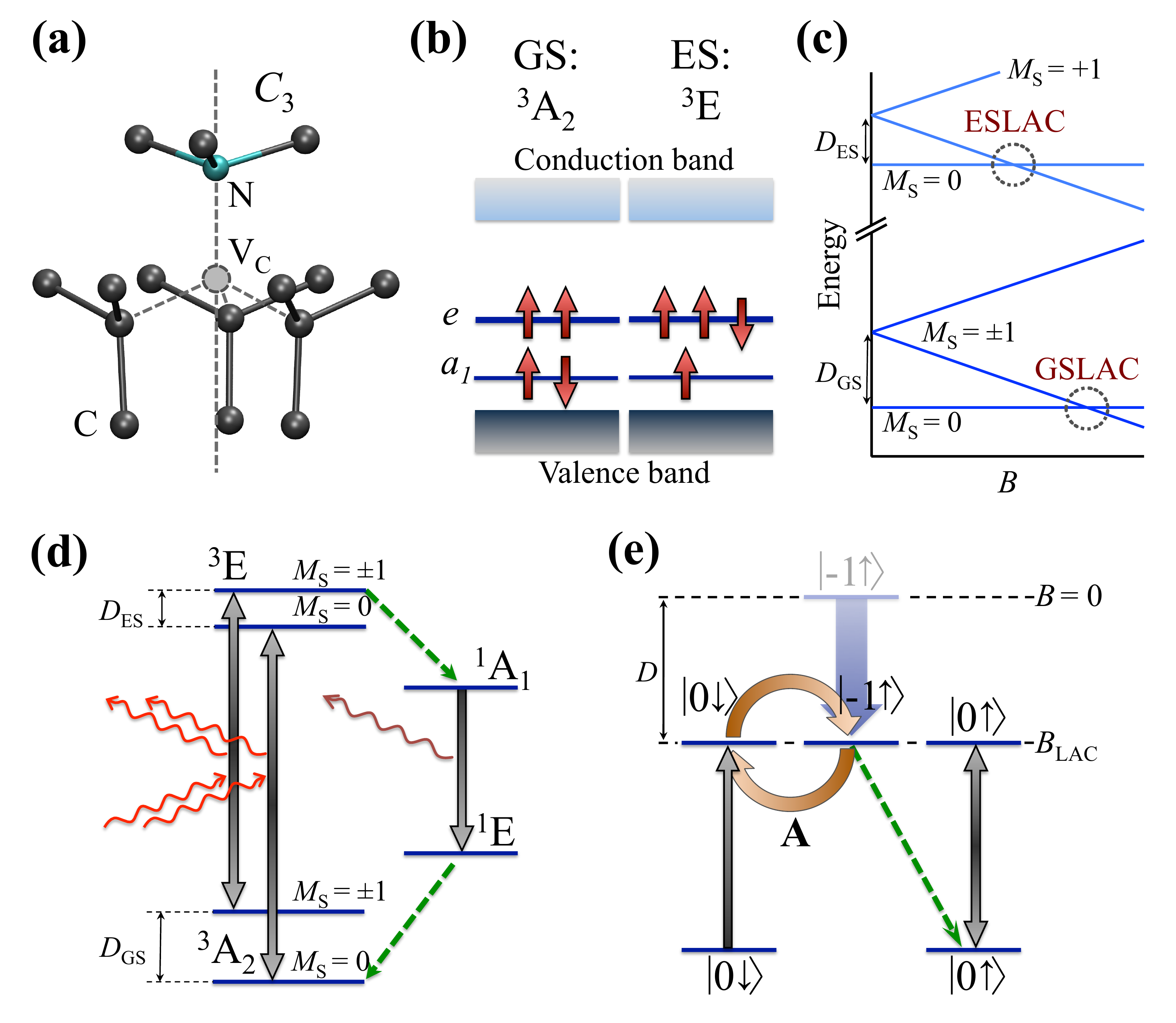}
\caption{\label{fig:NV} (Color on-line) A schematic diagram of (a) the structure showing the $C_3$ rotation axis and (b) the electron configuration of the NV center in diamond. (c) The magnetic field dependence of the ES and GS spin-sublevel energies, showing the ESLAC and GSLAC. (d) The NV center's optical polarization cycle and (e) the nuclear spin polarization cycle.  In (c) and (d), $D_\text{GS}$=2.87~GHz, $D_\text{ES}$=1.42~GHz are the zero-field constants. In (d), the green dashed arrows represent the non-radiative decays, which are mediated by spin-orbit couplings and vibrations. The thick grey arrows represent the optical absorption/emission paths. The wavy lines represent photon absorption/emission in the visible (red) and near-infrared (brown) regions. (e) At zero magnetic field ($B=0$), the $\left|0\downarrow\right\rangle$ level is separated by the zero-field constant from the $\left|-1\uparrow\right\rangle$ level. Applying a $B=B_\text{LAC}>0$ field causes the two states to form an avoided crossing, where the small gap is introduced by the hyperfine interaction ($\mathbf{A}$). In this condition, hyperfine coupling (brown circular arrow) and non-radiative decay (green dotted line) are responsible for the nuclear spin polarization in the optical cycle.}
\end{figure}

The geometry and the electronic structure of the negatively charged NV center in diamond have been discussed previously based on highly convergent \emph{ab initio} plane wave large supercell calculations \cite{Gali:PRB2008, Gali2009, Szasz2013}. The diamond NV center is a complex that consists of a substitutional nitrogen adjacent to a vacancy in diamond and possesses $C_{3v}$ symmetry (see Fig.~\ref{fig:NV}). The defect exhibits a fully occupied lower $a_1$ level, and a double-degenerate upper $e$ level filled by two parallel-spin electrons in the gap with an $S=1$ high-spin ground state. The $S=1$ excited state is well described by the promotion of an electron from the lower defect level to the upper level in the gap \cite{Gali:PRL2009}. The electron spin may interact with nuclear spins: $^{14}$N or $^{15}$N possessing $I=1$ or $I=1/2$ nuclear spin, respectively, or $^{13}$C with $I=1/2$.

The hyperfine interaction between the electron spin and nuclear spin has been studied by means of \emph{ab initio} methods in previous publications, both in the GS \cite{Gali:PRB2008, Smeltzer2011, Szasz2013,Nizovtsev2014} and in the ES \cite{Gali2009}. In the GS, the electronic spin-spin dipole interaction causes the fine electron structure to have a zero-field splitting, where ``zero field'' refers to zero external magnetic field. In high purity and low strain diamond samples, this splitting can be described by a single parameter, $D_\text{GS}$=2.87~GHz, which separates the $m_S=0$ and the $m_S=\pm 1$ sublevels within the $S=1$ manifold. At room temperature, the fine structure in the electronic ES shows a similar feature, except that its zero-field splitting is only $D_\text{ES}$=1.42~GHz. At lower temperatures the fine structure becomes more complicated, hindering off-resonantly pumped DNP in the ES \cite{Batalov:PRL2009}. Between the ES and GS triplets, non-radiative relaxation pathways through singlet states selectively flip $m_S=\pm 1$ states to $m_S=0$ state in the optical excitation cycle \cite{Harrison:DRM2006, Wrachtrup:JPCM2006, Manson:PRB2006, Maze2011, Doherty2011}, which allows optical spin-polarization of NV center (Fig.~\ref{fig:NV}).   
   
\begin{figure} [h!]
\includegraphics[width=0.9\columnwidth]{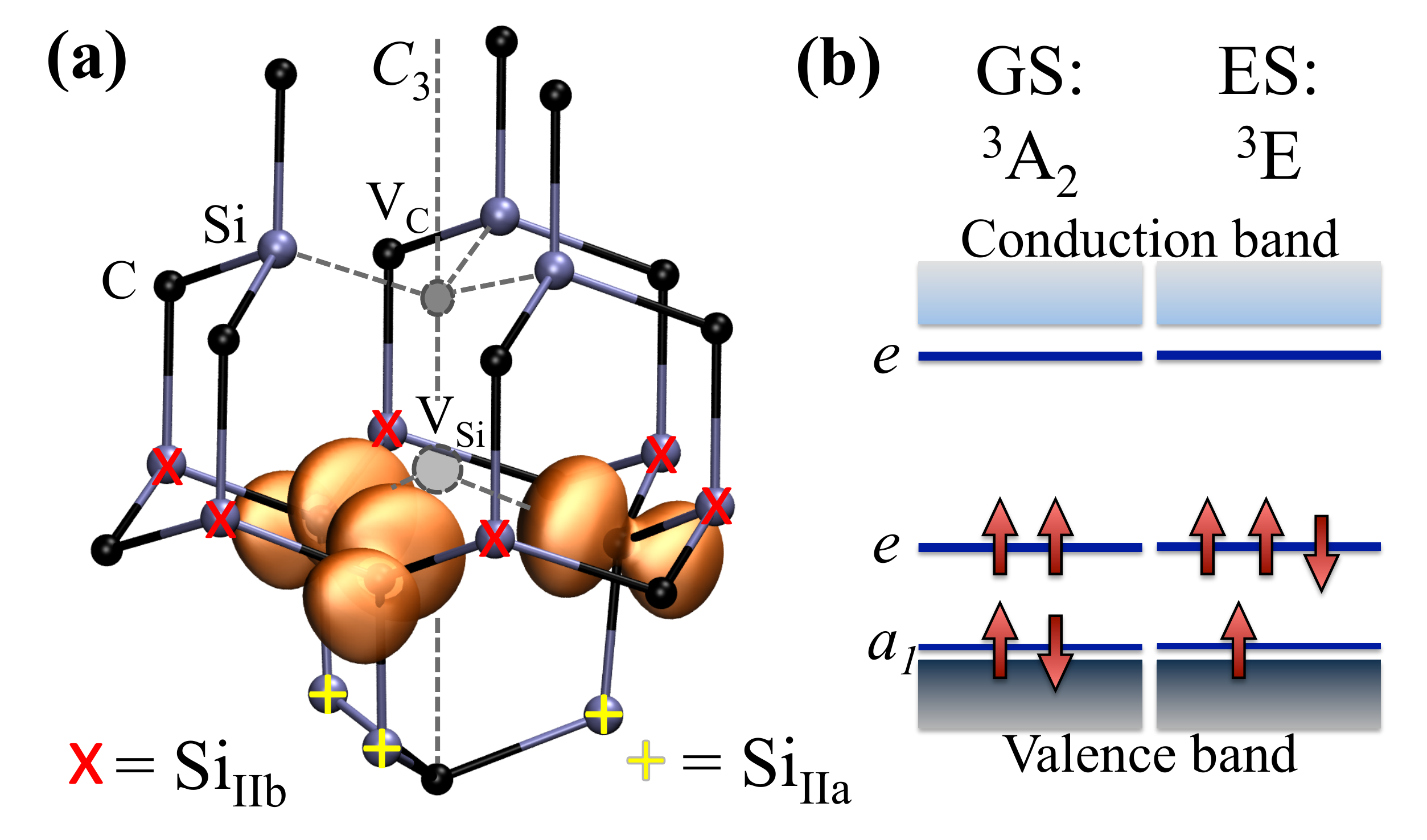}
\caption{\label{fig:V2} A schematic diagram of the structure of the divacancy in SiC. The defect levels in the gap as well as the corresponding ground and excited states are shown. The high energy upper empty $e$ level does not play a role in the excitation process. Those Si atoms that participate in DNP are labeled by Si$_\text{IIa}$ and Si$_\text{IIb}$. The similarity between the electronic structure of the NV center in diamond (c.f., Fig.~\ref{fig:NV}) and that of the SiC divacancy is apparent.}
\end{figure}

We also study the divacancy defects in SiC. Before describing the electronic structure of the divacancy, we will briefly discuss the host semiconductor. SiC has about 250 known different polytypes, which share the same basal hexagonal lattice but have different stacking sequences of Si-C bilayers perpendicular to this plane. The most industrially important polytypes are the 4H and 6H polytypes. The inequivalency of the crystal planes leads to the so-called $h$ and $k$ types of bilayers in 4H SiC, and $h$, $k_1$, $k_2$ types of bilayers in 6H SiC. In the inequivalent bilayers, the crystalline environment in the second, third, etc.\ neighborhood will be different, yielding a similar but quantitatively distinguishable electron structure for the corresponding point defects. For divacancy defects, adjacent silicon and carbon atoms are absent. In 4H SiC, the four inequivalent forms of divacancy are the $hh$, $kk$, $hk$ and $kh$ configurations. The first two (axial) configurations have $C_{3v}$ symmetry, and the second two (basal-plane oriented) configurations have $C_{1h}$ symmetry. We focus our study here on the axial configurations, whose $C_{3v}$  symmetry is the same as that of the NV center in diamond. In 6H-SiC, there are three axial configurations of divacancies: the $hh$, $k_1k_1$ and $k_2k_2$ configurations. In 4H-SiC, the $hh$ and $kk$ configurations have been already associated with zero-phonon-line photoluminescence peaks and spin transitions, measured with optically detected magnetic resonance (ODMR) signals and sometimes called the PL1 and PL2 centers \cite{Koehl11, Falk2014, Son2006}. An additional center has been found, PL6, which also exhibits ODMR and similar physical properties to the axial divacancies \cite{Koehl11,Falk2014,FalkPRL2015}. The physical structure of PL6 has not yet been identified. In 6H-SiC, the PL and ODMR lines that have been labeled QL1, QL2, and QL6\cite{Falk2013} are also associated with the axial divacancies (Section~\ref{sec:abinitio}), namely the $k_1k_1$, $hh$ and $k_2k_2$ configurations, respectively. All these axial divacancies share the same electronic structure depicted in Fig.~\ref{fig:V2}. The carbon dangling bonds of the Si-vacancy part of the defect create a double-degenerate $e$-level close the valence band edge, which is occupied by two electrons with parallel spins. Thus, the neutral divacancy has a high spin ($S=1$) ground state. The excited state may be described as the promotion of an electron from the lower defect $a_1$ level to this $e$ level \cite{Gali10, Gali11pss}, akin to that of the NV center in diamond. In SiC crystals, beside $^{13}$C isotopes (with a 1.3\% natural abundance), $^{29}$Si isotopes (with a 4.7\% natural abundance) have $I=1/2$ nuclear spins that may interact with the divacancies' $S=1$ electron spins.
Very little is known about the nature of the triplet ES and the dark singlet states responsible for the electron spin polarization for divacancy in SiC. However, recent measurements indicate \cite{FalkPRL2015} that the triplet ES has a similar electronic structure to that of NV center in diamond. Experimental results imply $C_{3v}$ symmetry in the ES of axial divacancies even at low temperatures \cite{FalkPRL2015}. In this work, we assume that the SiC divacancy has a similar model for the optical spin polarization processes as does the NV center in diamond. The measured $D_\text{GS}$ and $D_\text{ES}$ 
zero-field splittings of axial divacancies in 4H and 6H polytypes are summarized in Table~\ref{tab:D}.

\section{Model of the dynamic nuclear spin polarization process}
\label{sec:method}

\subsection{Modeling the dynamic nuclear spin polarization in the optical cycle}

The degree of nuclear spin polarization ($P$) is the main observable in DNP measurements. Understanding its dependence on external variables like magnetic field and temperature, and internal variables like the details of the hyperfine tensor, has great importance from an applications point of view. A theoretical model is a crucial component of this understanding.

In the forthcoming section, we briefly review the model of Jacques \emph{et al}.\ \cite{Jacques2009} on the dynamic nuclear spin polarization of the NV center's nitrogen nucleus and extend it by a generalized derivation of the basic equations. Next, we describe our model that takes into account many microscopic features and processes that have been overlooked or not integrated in previous models. The most important features that we integrate are the electron spin decoherence in the ground and excited state, the short lifetime of the excited state, the anisotropy of the hyperfine tensors, the angle of the external magnetic field, and the overlap of the ground and excited state's spin flipping processes.

\subsubsection{Previous model on dynamic nuclear spin polarization}

The dynamic nuclear spin polarization processes are sensitive to the details of the spin Hamiltonian of the considered point defect and the optical electron spin polarization cycle. In the simplest case, e.g.\ an NV center in diamond and an adjacent $I=1/2$ nuclear spin locate on the symmetry axis of the defect, the process can be understood as a two-step mechanism. Continuous optical excitation polarizes the electron spin in the $M_{S}=0$ spin state, while the nuclear spin may be any linear combination of the spin-up $\left| \uparrow \right\rangle$ and spin-down $\left| \downarrow \right\rangle$ states. Near the vicinity of a LAC, the hyperfine interaction couples the nuclear and electron spins effectively and rotates the two spin state $\left| 0  \downarrow\right\rangle$ into the state $\left| -1 \uparrow \right\rangle$. This state is then transformed into the state $\left| 0 \uparrow \right\rangle$ by the optical excitation and decay processes shown in Fig.~\ref{fig:NV} (e). As a result, the nuclear spin component $\left| \downarrow \right\rangle$ is flipped and the $\left| \uparrow \right\rangle$ is predominantly populated.

The spin Hamiltonian that governs the nuclear spin flips is well known for the NV-center in diamond \cite{Jacques2009}. For a single nuclear spin adjacent to the defect, the ground-state Hamiltonian of the electron-nuclear spin system can be written as
\begin{equation} \label{eq:GS_Ham}
\hat{H}_{\text{GS}}= \hat{\mathbf{S}}^{\text{T}} \mathbf{D}_{\text{GS}} \hat{\mathbf{S}} + \mu_{\text{B}} \mathbf{B}^{\text{T}} \mathbf{g}_{\text{e}} \hat{\mathbf{S}} + \hat{\mathbf{S}}^{\text{T}} \mathbf{A}_{\text{GS}} \hat{\mathbf{I}} + \mu_{\text{N}} \mathbf{B}^{\text{T}}  \mathbf{g}_{\text{N}} \mathbf{\hat{I}}\text{,}
\end{equation}
where $\hat{\mathbf{S}}$ and $\hat{\mathbf{I}}$ are the electron and nuclear spin operators, $\mathbf{D}_{\text{GS}}$ and $\mathbf{A}_{\text{GS}}$ are the tensors of zero-field and hyperfine interaction in the ground state electron spin configuration of the defect, respectively, $\mathbf{B}$ is the external magnetic field, $\mathbf{g}_{\text{e}}$ and $\mathbf{g}_{\text{N}}$ are the $g$ tensors of the electron and nuclear spin, and $\mu_{\text{B}}$ and $\mu_{\text{N}}$ are the Bohr and nuclear magnetons, respectively. For the considered defect, the $g$ tensors are nearly isotropic and simplify to a scalar value, such as $g_{\text{e}} = 2.0023$.

For the NV center, the zero-field interaction term $\hat{H}_{\text{zfs}}$, the first term on left hand side of Eq.~(\ref{eq:GS_Ham}), can be written as
\begin{equation}
\hat{H}_{\text{zfs}}  = \hat{\mathbf{S}}^{\text{T}} \mathbf{D}_{\text{GS}} \hat{\mathbf{S}} = D_{\text{GS}} \left( \hat{S}^{2}_{z} - \frac{2}{3} \right) \text{.}
\end{equation}

The hyperfine interaction term $\hat{H}_{\text{hyp}}$, which is the third term on the left hand side of Eq.~(\ref{eq:GS_Ham}), couples the electron and nuclear spins and therefore plays a key role in DNP. When the nuclear spin resides on the symmetry axis of the defect the hyperfine tensor $\mathbf{A}$ is diagonal with diagonal elements $A_{\perp}$ and $A_{\parallel}$. The hyperfine interaction term is written as\cite{Gali2009}
\begin{equation} \label{eq:Hyp_ax}
\hat{H}_{\text{hyp}} = \hat{\mathbf{S}}^{\text{T}} \mathbf{A}_{\text{GS}} \hat{\mathbf{I}} = A_{\perp}^{\text{GS}} \frac{\hat{S}_{+}\hat{I}_{-} + \hat{S}_{-}\hat{I}_{+}}{2} + A_{\parallel}^{\text{GS}} \hat{S}_{z}\hat{I}_{z} \text{,}
\end{equation}  
where $\hat{S}_{\pm}$ and $\hat{I}_{\pm}$ are the electron and nuclear spin ladder operators, respectively, and $\hat{S}_{z}$ and $\hat{I}_{z}$ are the $z$ components of the electron and nuclear spins, respectively. The first term on the left hand side of Eq.~(\ref{eq:Hyp_ax}) is responsible for the flipping of the nuclear and electron spins governed by $A_{\perp}$.

The NV-center in diamond possesses $C_{3v}$ symmetry in its excited state (ES) at elevated temperatures \cite{Gali:PRB2008}, and therefore the excited state Hamiltonian has the same form as Eq.~(\ref{eq:GS_Ham}). However, the zero-field-splitting tensor $ \mathbf{D}_{\text{ES}}$ and the hyperfine tensor $\mathbf{A}_{\text{ES}}$ differ from those in the ground-state electronic configuration. 

For the sake of a general description, we utilize the density matrix formalism\cite{Perel1973} to derive the steady state nuclear spin polarization of the dynamical process. First, we express the wave function of an electron and nuclear spin system in a basis, 
\begin{equation}
\Psi = \sum_{m,n} C_{m n} \! \left( t\right) \left| m n \right\rangle \text{,}
\end{equation}   
where $\left | m n \right\rangle = \left| m\right\rangle \otimes \left| n \right\rangle$,  $m$ and $n$ are the electron and nuclear spin projections on the quantization axis and $C_{m n} $ are coefficients. The spin-density matrix of the system can be obtained from the coefficients,
\begin{equation}
\rho_{m n, {m}' {n}'} \! \left( t \right) = C_{m n} \! \left( t \right) C^{*}_{m' {n'}} \! \left( t \right)  \text{,}
\end{equation}
while the nuclear spin-density matrix can be obtained from the partial trace of $\rho_{m n, {m}' {n}'}$, 
\begin{equation}
\Phi_{n {n}'} \! \left( t \right) = \sum_{m} \rho_{m n, m {n}'} \! \left( t \right) \text{.}
\end{equation}
To describe the time evolution of the diagonal elements of the nuclear spin density matrix in the dynamical process, for timescales larger than the average length of a dynamical cycle, we write the kinetic equations in the form
\begin{equation}
\dot{\Phi}_{n n} = c \! \left( J \right) \overline{ \Delta \Phi}_{n n} - \eta \Phi_{n n} \text{,}
\end{equation} 
where the dot represents time differentiation. The second term on the right hand side describes the nuclear spin relaxation due the environment. Note that all the nuclear spin projections are assumed to relax equally with rate $\eta$. The first term on the right hand side describes the change of the nuclear spin projection due to the interaction with the electron spin and the external magnetic field. $c \! \left( J \right) $ is the number of optical cycle per unit time and $\overline{ \Delta \Phi}_{n n}$ is the averaged variation of the nuclear spin projection during the free evolution time, i.e.\ between two optical excitations. This last term can be determined from the spin Hamiltonian of the system. Note that both terms depend on the intensity $J$ of the excitation laser. At low intensities, the rate of optical excitation $c \! \left( J\right)$ depends linearly on the intensity. For strong laser excitation, when the rate of excitation may be comparable with the periodicity of the spin rotations due to hyperfine interaction or transverse magnetic field, the time average of $ \Delta \Phi_{n n}$ depends on $J$ too. In the following, we assume weak laser intensities, and thus only $c\left( J\right)$ depends on the intensity. Generally, $\overline{ \Delta \Phi}_{n n}$ can be written as
\begin{equation}\label{eq:adndm}
\overline{ \Delta \Phi}_{n n} = \sum_{{n}'} \bar{p}_{{n}' n} \Phi_{{n}'{n}'} - \Phi_{n n} \sum_{{n}'} \bar{p}_{n {n}'} \text{,}
\end{equation}
where $\bar{p}_{n {n}'}$ is the average probability of flipping the nuclear spin from $\left| n \right\rangle$ to $\left| {n}' \right\rangle$.
The first and second terms on the right hand side describe the probabilities of flipping the spin in and out of the state $\left| n \right\rangle$, respectively. We note that for the Hamiltonian specified in Eq.~(\ref{eq:GS_Ham})-(\ref{eq:Hyp_ax}), $n' = n \pm 1$ holds\cite{Kalevich2015}. In the following, we restrict our derivation to the case of $I=1/2$, however, we do not impose any other losses in the generality of the spin Hamiltonian. In this case, Eq.~(\ref{eq:adndm}) reads as
\begin{equation}
\overline{ \Delta \Phi}_{\pm\frac{1}{2}\pm\frac{1}{2}} = p_{\pm} \Phi_{\mp\frac{1}{2}\mp\frac{1}{2}} - p_{\mp} \Phi_{\pm\frac{1}{2}\pm\frac{1}{2}} \text{,}
\end{equation}  
where $p_{\pm}$ are the average probabilities of raising and lowering of the nuclear spin projection between two optical cycles. 

When dynamic nuclear spin polarization is in a stationary state, the different spin rotation processes are balanced and:
\begin{equation} \label{eq:eq}
\dot{\Phi}_{+\frac{1}{2}+\frac{1}{2}} = \dot{\Phi}_{-\frac{1}{2}-\frac{1}{2}} \text{.}
\end{equation}
By using the definition $P = \Phi_{+\frac{1}{2}+\frac{1}{2}} - \Phi_{-\frac{1}{2}-\frac{1}{2}}$ and the normalization condition  $\Phi_{+\frac{1}{2}+\frac{1}{2}} + \Phi_{-\frac{1}{2}-\frac{1}{2}} = 1$, from Eq.~(\ref{eq:eq}) we can express the steady state nuclear spin polarization as
\begin{equation} \label{eq:pol}
P = \frac{p_{+} - p_{-}}{p_{+} + p_{-} + \kappa}\text{,}
\end{equation}
where $\kappa \equiv  \eta / c \left(  J \right)$.

As can be seen, the flipping probabilities $p_{+} $ and $p_{-}$ play an important role in this process. In the previous model\cite{Jacques2009}, these were determined analytically from the simplified spin Hamiltonian Eq.~(\ref{eq:GS_Ham})-(\ref{eq:Hyp_ax}) for the case $S = 1$ and $I = 1/2$, as
\begin{equation}
p_{+} \! \left( B\right) = 2 \left| \left\langle \left. 0 \downarrow \right| + \right\rangle  \right|^2_{B} \left| \left\langle \left. -1 \uparrow \right| + \right\rangle  \right|^2_{B} \text{,}
\end{equation}
\begin{equation}
p_{-} \! \left( B \right) = p_{+} \! \left( -B\right) \text{,}
\end{equation}
where $\left| + \right\rangle$  is the eigenstate of $\hat{H}_{\text{ES}} \! \left( B\right)$. This state is a mixture of $\left| 0 \downarrow \right\rangle$ and $\left| -1 \uparrow\right\rangle$ states due to hyperfine coupling.

\subsubsection{Our extended model}

We now discuss the details of our extended model, which is generally applicable and shows predictive power.  First, we apply modifications to the spin Hamiltonian in Eq.~(\ref{eq:GS_Ham}).

In previous models, the external magnetic field was aligned parallel to the $C_{3v}$ axis of the defect. However, experiments show that DNP strongly depends on the misalignment of the magnetic field\cite{Jacques2009}. To describe the effect of this misalignment, we allow small deviations of the direction of the magnetic field from the symmetry axis of the defects. Furthermore, the nuclear Zeeman effect was also neglected in previous models. At stronger magnetic fields, such as at 500-1000~Gauss, the nuclear Zeeman splitting can reach few MHz, which can be comparable to the hyperfine interaction. We thus take this effect into account in our model.     

In the general case, the nucleus with non-zero spin is not on the symmetry axis of the defect. In this case, the symmetry of the spin Hamiltonian is reduced, i.e.\ the hyperfine tensor $\mathbf{A}$ may have three non-degenerate eigenvalues $A_{xx}$, $A_{yy}$, and $A_{zz}$, and the eigenvector $\mathbf{a}_z$, which corresponds to the eigenvalue $A_{zz}$, may have a non-zero angle of $\theta$ with the symmetry axis. The azimuthal angle $\varphi$ may be chosen to zero without limiting the generality. The effects of symmetry-breaking hyperfine interactions have been included in previous considerations to some extent\cite{Smeltzer2009,Gali2009}. However, a consistent description of the DNP process with a general hyperfine tensor has not been carried out so far. 

In our model, we consider non-diagonal hyperfine tensors, which can be parameterized by their eigenvalues and angle $\theta$ as follows:
\begin{equation}
\hat{H}_{\text{hyp}}  = \hat{\mathbf{S}}^{\text{T}}  \mathbf{A} \hat{\mathbf{I}}  =  \left( \mathbf{U} \hat{\mathbf{S}} \right)^{\text{T}}  \!\! \mathbf{A}_{\text{diag}}  \!\! \left( \mathbf{U} \hat{\mathbf{I}} \right) \text{,}
\end{equation}  
where $\mathbf{U}$ describes a rotation that transforms the Cartesian basis to the eigenbasis of tensor $\mathbf{A}$, and  $\mathbf{A}_{\text{diag}} = \mathbf{U} \mathbf{A} \mathbf{U}^{\text{T}}$ is the diagonal tensor of elements $A_{xx}$, $A_{yy}$, and $A_{zz}$. Note that for a general hyperfine tensor the spin Hamiltonian may contain $\hat{S}_{\pm} \hat{I}_{z}$, $\hat{S}_{z} \hat{I}_{\pm}$, and $\hat{S}_{\pm} \hat{I}_{\pm}$ terms that allow a wide range of spin rotation processes to occur (see Appendix for spin Hamiltonian matrices). 

The nuclear spin flipping probabilities play a key role in the determination of the polarization, defined in Eq.~\ref{eq:pol}. The main innovation of our model is that it uses different definitions for the probabilities $p_{+}$ and $p_{-}$ than previous models\cite{Jacques2009} and includes important effects from the excited and ground states' spin Hamiltonians as well as external driving forces from the surrounding spin bath. In the rest of this section, we describe the main concept of our new considerations.

For simplicity, in this section we restrict ourselves to the case of positive external magnetic field, which shifts the energies of $M_{S}=+1$ and $M_{S}=-1$ levels upward and downward, respectively. In such a case, at vicinity of $B_{\text{LAC}}$, the $M_{S}=-1$ spin state mixes with the $M_{S}=0$ state. For simplicity, we do not consider the interaction of the $M_{\text{S}}=+1$ state with other states here. However, we include it in our later calculations (see Section~\ref{sec:method_prob}). 

Nuclear spin rotation can occur both in the electronic ground and excited states. The interplay of these rotations and non-radiative, spin-selective electronic decay is responsible for the nuclear spin polarization. In the most general case, to achieve a net driving force toward nuclear spin polarization, the ground-state and the subsequent excited-state spin rotation processes have to fulfill the following criterion: starting from the $M_{S}=0$ electron spin state, the electron spin \emph{may be flipped} into the $M_{S}=-1$ state, but the starting nuclear spin state \emph{must be flipped} into the opposite spin projection. When electron spin flip-flops occur, the $M_{S} = -1$ electron spin is then transported into the initial $M_{S}=0$ state by non-radiative decay. Optical cycles can therefore flip nuclear spins. When the rates of the flipping processes $\left| 0 \downarrow \right\rangle \rightarrow \left| 0 \uparrow\right\rangle$  and $\left| 0 \uparrow \right\rangle \rightarrow \left| 0 \downarrow \right\rangle$ are different due to some sort of asymmetry, the repetition of these dynamical cycles induces different population of the nuclear spin states and non-zero polarization.

\begin{figure}
\includegraphics[width=0.9\columnwidth]{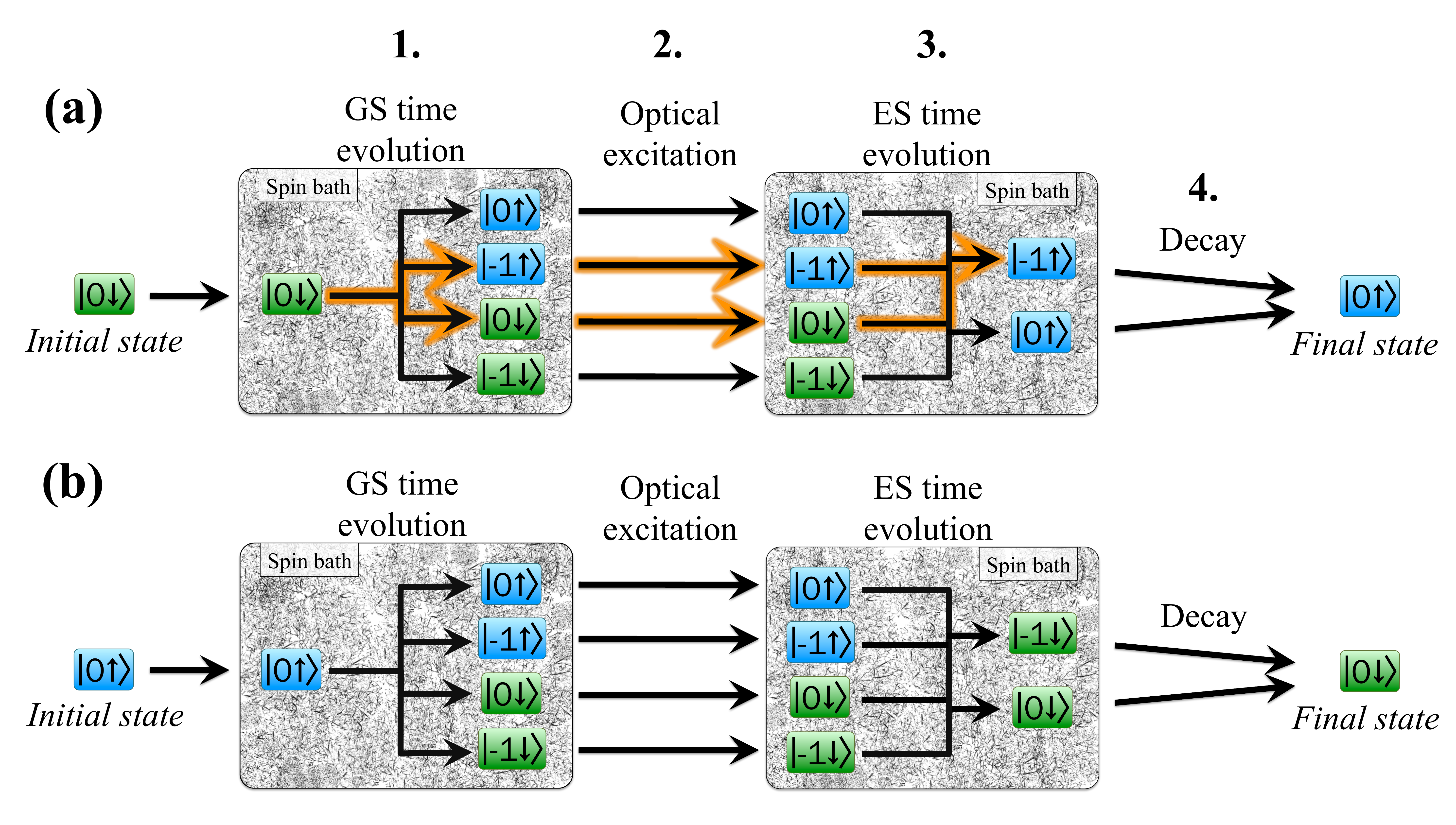}
\caption{\label{fig:cycle}(Color online)  A schematic diagram of the evolution of a coupled electron and nuclear spin system through the four steps of a complete dynamic nuclear spin polarization cycle. The depicted spin rotation processes, shown by different paths along the black arrows, preserve the spin projection $M_{S}=0$ of the electron spin but flip the nuclear spin (a) from down to up, and (b) from up to down, in a four-step mechanism. The states of up and down nuclear spin projections are depicted with blue and green backgrounds, respectively.  In the general case, the interplay of all of these parallel processes determines the net polarization of the nuclear spins, with the process rates defined by the ground and excited state spin Hamiltonians (see text and Appendix for more information).  The upper and lower path of arrows with orange (thick light grey) outline show the most prominent spin rotation processes at the GSLAC and ESLAC for positive magnetic fields, respectively. The background spin bath, which works against DNP, is represented by grey background.}  
\end{figure}

All the possible spin rotation processes that fulfill the aforementioned criterion and are allowed by the general spin Hamiltonian (see Appendix for details) are depicted on a schematic model of the dynamical cycle in Fig.~\ref{fig:cycle}. The probability of the ground or excited state spin rotation processes $p^{\text{St}}\!\left(  \left. \chi_{\text{Initial}} \right| \chi_{\text{Final}} \right)$, where $\chi_{\text{Initial}} $ and $\chi_{\text{Final}}$ represents the initial and final spin configurations, respectively, can be determined by the spin Hamiltonian of the defect in the considered states 'St'. The probability of spin flip in a joint ground state - excited state spin rotation process is then the product of the probabilities of the two separate rotations. With the above requirement, the products $p^{\text{GS}}\!\left(  \left. 0, \pm 1/2 \right| \chi_{\text{Inter}} \right) p^{\text{ES}}\!\left(  \left. \chi_{\text{Inter}} \right| -1, \mp1/2 \right)$ and $p^{\text{GS}}\!\left(  \left. 0, \pm 1/2 \right| \chi_{\text{Inter}} \right) p^{\text{ES}}\!\left(  \left. \chi_{\text{Inter}} \right| 0, \mp1/2 \right)$   define the probability of spin rotations that induce driving forces toward nuclear spin polarization, where $\chi_{\text{Inter}}$ represents an intermediate spin configuration and $\left | \pm 1/2 \right \rangle$ is a comprehensive notation for $\left | \uparrow \right \rangle$ or $\left | \downarrow \right \rangle$ nuclear spin states.

In contrast to previous models, our two-state evaluation model requires suitable spin rotation mechanism for DNP not separately but simultaneously in the ground and excited states. For example, at $B_\text{GSLAC}$, the ground-state hyperfine coupling flips the nuclear and electron spins with high probability. This process is represented by orange arrow on the top panels of Fig.~\ref{fig:cycle}. Optical excitation transports this spin state to the excited state, where it starts to evolve in accordance with the new spin Hamiltonian of the excited state. To obtain a high probability for suitable electron and nuclear spin flips from the two successive time evaluations, the spin state must be unchanged in the excited state. For general hyperfine tensors of the excited and ground state that are not necessarily identical, this condition may not be the case. The role of minority processes, represented with black arrows in Fig.~\ref{fig:cycle}, increases as the symmetry of the system becomes more distorted. Examples of this sort of distortion include a misaligned magnetic field or hyperfine interactions with low symmetry.  Such effects can lower the polarizability of the system or cause unexpected resonance effects, which can be captured by our two-state evaluation model.

The next new consideration in our model is the determination of the preferential direction of the nuclear spin polarization. The DNP process is in a stationary state when the polarization and depolarization processes are in equilibrium (i.e.\ Eq.~(\ref{eq:eq}) is satisfied). This state corresponds to a certain electron and nuclear spin configuration, which is not necessarily equal to one of the energy eigenstates of the ground or excited-state spin Hamiltonian. Generally, the nuclear spin is polarized or preserved in a state that has the longest lifetime in the dynamical cycle. When the hyperfine interaction is isotropic or $C_{3v}$ symmetric, the energy eigenstate $\left| 0 \uparrow \right\rangle$ has the longest, quasi-infinite lifetime, since in this state there is no hyperfine coupling between the electronic and nuclear subsystems\cite{Jacques2009}. On the other hand, for general hyperfine interactions of the excited and ground state, the longest-lived spin state, which is the stationary spin state, can be a non-symmetric state\cite{Fischer2013}.

This non-symmetric state can be written as a linear combination $ \left| 0  \right\rangle \otimes  \left( \alpha \left| \uparrow \right\rangle + \beta \left| \downarrow \right\rangle \right)$. If the nuclear spin state is the eigenstate of the nuclear spin operator $\hat{I}_{\mathbf{e}}$ of quantization axis $\mathbf{e}$, then unit vector $\mathbf{e}$ of the three dimensional space shows the preferential direction for nuclear spin polarization in the dynamical cycle. In our model, we determine this $\mathbf{e}$ direction for every different set of parameters. To find this direction, we used the following condition: the nuclear spin state $\left| \uparrow_{\mathbf{e}} \right\rangle$, which is the eigenstate of $\hat{I}_{\mathbf{e}}$ with eigenvalue $+1/2$, has the longest lifetime when its precession is minimal. The precession would result in a continuous transition between the two eigenstates of $\hat{I}_{\mathbf{e}}$, i.e.\ $\left| \uparrow_{\mathbf{e}} \right\rangle  \leftrightarrow \left| \downarrow_{\mathbf{e}} \right\rangle$, which then reduces the lifetime of the nuclear spin state and lowers polarization. Since the system spends most of the time in the ground state during the dynamical cycle, we assumed that the ground-state precession should be minimal in the stationary state of the dynamical cycle. This means that we minimized $p^{\text{GS}} \!\! \left( \left. 0, \uparrow_{\mathbf{e}} \right| 0, \downarrow_{\mathbf{e}} \right)$ with respect to direction $\mathbf{e}$. At the minimum, we find the direction where the nuclear spin state $\left| \uparrow_{\mathbf{e}} \right\rangle$ is preserved for the longest time. This state is therefore the stationary nuclear spin state in the dynamical cycle.      

In the optical cycle, the length of the free evolution of the spin system is limited by the lifetime and decoherence of the electron states (see below). When the lifetime of the ground or excited state is shorter than the characteristic time of the spin rotation processes, the oscillatory probability is not averaged out. In extreme cases, the spin-flip probability is effectively reduced by the short lifetime of the electron state (see Fig.~\ref{fig:NV-tev} for the case of the NV center in diamond in Section~\ref{sec:NV}). This case is more pronounced in the excited state, since the excited state lifetime of both the NV center in diamond and the divacancy in SiC is only 10-15 ns. For a hyperfine interaction of 10~MHz, the oscillation time of the nuclear spin-flip probability is on the order of 100~ns at $B_{\text{LAC}}$, is much longer than the lifetime of the excited state. In contrast to previous models, we calculate the probabilities by explicitly taking time into account in the spins' excited-state evolution (see Section~\ref{sec:method_prob} for more detail). We assume that the evolution time in the excited state is exponentially decaying. The characteristic time of decay is connected to the lifetime of the excited state ($\tau_{\text{ES}}$) determined by experiment. On the other hand, we assume that the oscillatory probabilities are averaged out in the ground state, where the system spends sufficient time for this to happen. 

Finally, we include additional important effects in our model: electron spin dephasing and spin-lattice relaxation effects. The surrounding spin bath of nuclear spins and paramagnetic defects disturbs not only the nuclear but the electron spin of the considered spin system\cite{Fischer2013}, causing decoherence of the superposition states. This decoherence can be described by the Lindblad equation,
\begin{equation} \label{eq:Lindblad}
 \frac{\partial \hat{\rho}}{\partial t} = - \frac{i}{\hbar} \left[ \hat{H}, \hat{\rho} \right] + \hat{L} \hat{\rho} \text{,}
\end{equation}
where $\hat{\rho}$ and $\hat{H}$ are the density operator and the Hamiltonian of the considered system, respectively. The final term on the right hand side of Eq.~(\ref{eq:Lindblad}), which describes the effects of the environment, can be written as
\begin{equation}
\hat{L} \hat{\rho} = \sum_{n} \gamma_n \left( \hat{C}_n \hat{\rho} \hat{C}_n^{\dagger} - \frac{1}{2} \left\lbrace \hat{C}_n \hat{C}_n^{\dagger}, \hat{\rho} \right\rbrace \right) \text{,}
\end{equation}
where $\gamma_n$ are the decay rates of processes described by the Lindblad operators $\hat{C}_n$. Decoupling of the superposition state of the $\left| i \right\rangle$ and $\left| j \right\rangle$ states can be taken into account by the operator $\hat{C}= \left| i \right\rangle \!\! \left\langle i \right| - \left| j \right\rangle \!\! \left\langle j \right|$. In our case, the state $\left| + \right\rangle = \alpha \left| 0\downarrow \right\rangle + \beta \left| -1 \uparrow \right\rangle$, which is responsible for the most prominent nuclear spin flipping process, relaxes with $T_{2}^{*}$ characteristic time due to the electron spin dephasing processes can be accounted by the Lindblad operator $\left| 0 \right\rangle \!\! \left\langle 0 \right| - \left| -1 \right\rangle \!\! \left\langle -1 \right|$. Besides dephasing, electron spin-lattice relaxation can also hinder the nuclear spin flipping processes, by depopulating the hyperfine coupled states $\left| 0\downarrow \right\rangle $ and $ \left| -1 \uparrow \right\rangle$. This effect can be described by the Lindblad operators $\left| i \right\rangle  \!\! \left\langle  j \right|$.  Recent experiments have shown \cite{Fuchs2010} that the electron spin coherence time $T_{2}^{*}$ in the excited state of the NV center in diamond is on the order of the excited-state lifetime. To take into account the aforementioned effects, we assume that the intact evolution time of the electron and nuclear spin system, described by the spin Hamiltonian in Eq.~(\ref{eq:GS_Ham}), is effectively reduced in the excited state due to the short coherence time of the electron spin. Therefore, we used a scaled excited state lifetime $\tau_{ES}^{*} = \nu \tau_{\text{ES}}$ in our model, where $\nu$ is a free parameter of the property $0 < \nu < 1$.  By considering electron spin decoherence and spin-lattice relaxation effects,  $1 / \tau_{ES}^{*} \approx 1/ T_{2}^{*} + 1 / T_{1} $, where $T_{1} $ is the characteristic time of the spin-lattice relaxation.  As dephasing is the fastest spin relaxation process, $\tau_{ES}^{*} $ is assumed to be close to, but somewhat smaller than $T_{2}^{*}$. For the ground state, where the evaluation time is not included explicitly, we scaled down the probability of nuclear spin rotation processes by a factor of $\mu$.

Using the results of experiment and first-principles calculations for the parameters of the above described model, three free parameters remain. All of them are related to some extent to the effect of decoherence and spin-lattice relaxation. Parameter $\kappa \equiv \eta / c\left( J\right)$ in Eq.~(\ref{eq:pol}) is mainly related to the nuclear spin-lattice relaxation, and the two parameters $\nu$ and $\mu$, introduced above, are closely connected to the electron dephasing effects in the excited and ground state, respectively.

Despite these new considerations, the model that we have described still has limitations. Since it is a single cycle model, i.e.\ evolution of the spin system is taken into account in a single optical cycle, complicated processes that take place over many cycles are not modeled. Such a mechanism appears for $I \geq 1$ \cite{Smeltzer2009,Fischer2013}.

\subsection{Method of calculation of the nuclear spin polarization}
\label{sec:method_prob}

In this section, we specify the equations that are used to calculate the nuclear polarization, $P$, in the framework of our model described in the previous section.

As Eq.~(\ref{eq:pol}) shows, the polarization can be calculated from the probability of nuclear spin up and down flips ($p_{+}$ and $p_{-}$) in a dynamical cycle, respectively, and from the rate of spontaneous nuclear spin flips $\kappa$ due to the background spin bath. The latter quantity is one of our model's free parameters. Since the LAC happens for two values of the external magnetic field, $\pm B_{\text{LAC}}$, the nuclear spin flipping probabilities can be divided up into two parts,
\begin{align}
p_{+}  &= p_{+}^{\left(-1\right)} + p_{+}^{\left(+1\right)}, \\ \nonumber
p_{-}  &= p_{-}^{\left(-1\right)} + p_{-}^{\left(+1\right)}\text{,}
\end{align}
where $p_{+}^{\left(-1\right)} $ and  $p_{-}^{\left(-1\right)} $ and $p_{+}^{\left(+1\right)} $ and  $p_{-}^{\left(+1\right)} $ represent the spin up and down flipping probabilities due spin rotation processes in the subspace $M_{\text{S}}=\left\lbrace 0, -1 \right\rbrace$ and $M_{\text{S}}= \left\lbrace 0,+1\right\rbrace$ , respectively. For positive values of $B$ the second terms on the right hand side of the equations have only minor contribution. However, their role increases as $B \rightarrow 0$. At $B=0$, $p_{+}$ and $p_{-}$ become equal, and therefore, $\left. P \right|_{B=0} = 0$. In our model, we use the approximation\cite{Jacques2009} $p_{+}^{\left(+1\right)} \! \left( B \right) = p_{-}^{\left(-1\right)} \! \left( -B \right)$ and $p_{-}^{\left(+1\right)} = p_{+}^{\left(-1\right)} \! \left( -B\right)$. The nuclear spin-flipping probabilities, corresponding to the $M_{\text{S}} = \left\lbrace 0, -1 \right\rbrace$ subspace, can be defined as,
\begin{equation}\label{eq:p_pm}
p_{+}^{\left(-1\right)} =\sum_{i} p^{\text{GS}} \! \left( \left. 0\downarrow \right| \chi_{i} \right) \left[  p^{\text{ES}} \! \left( \left. \chi_{i} \right| -1 \uparrow \right)  \Gamma  + p^{\text{ES}} \! \left( \left. \chi_{i} \right| 0 \uparrow \right)   \right]\text{,}
\end{equation} 
\begin{equation} \nonumber
p_{-}^{\left(-1\right)} = \sum_{i} p^{\text{GS}} \! \left( \left. 0 \uparrow \right| \chi_{i} \right)  \left[ p^{\text{ES}} \! \left( \left. \chi_{i} \right| -1 \downarrow \right)  \Gamma + p^{\text{ES}} \! \left( \left. \chi_{i} \right| 0 \downarrow \right)  \right]\text{,}
\end{equation}
where $\Gamma$ is the probability of non-radiative decay from the electron spin state $\left| \pm 1\right\rangle$ of the excited state to the ground state spin state $\left| 0\right\rangle$ and $\chi_i$ are the final and initial states of the ground and excited states' time evolution, respectively. The summation goes over states $\left|0 \uparrow \right\rangle$,$\left|0 \downarrow \right\rangle$, $\left| -1 \uparrow \right\rangle$, and $\left|-1 \downarrow \right\rangle$, see Fig.~\ref{fig:cycle}.

To evaluate Eq.~(\ref{eq:p_pm}), we define the probabilities $p^{\text{St}}\!\left(  \left. \chi_{\text{Initial}} \right| \chi_{\text{Final}} \right)$. For the ground state spin rotation processes, we averaged out $ \left| \left\langle \left. \chi_{\text{Final}} \right| \chi\! \left( t \right) \right\rangle \right|^{2}$ in time, as
\begin{equation} \label{eq:p_gs}
p^{\text{GS}} \! \left(  \left. \chi_{\text{Initial}} \right| \chi_{\text{Final}} \right) = \sigma^{\text{GS}}_{\text{I},\text{F}} \frac{1}{T_{\text{GS}}} \int^{T_{\text{GS}}}_{0}  \left| \left\langle \chi_{\text{Final}} \right| e^{-i \hat{H}^{\text{GS}} t / \hbar} \left| \chi_{\text{Initial}} \right\rangle \right|^{2} dt \text{,}
\end{equation}
where we consider the integration time $T_{\text{GS}}\rightarrow \infty$. This limit means that the system spends enough time in the ground state for the oscillatory probabilities to be completely averaged out. The factor $\sigma^{\text{GS}}_{\text{I},\text{F}}$ takes into account the destructive effect of the electron spin decoherence and spin-lattice relaxation. $\sigma^\text{GS}_{\text{I},\text{F}}$ is a two-value function that takes $1$ when the initial and final states are the same and takes $0<\mu<1$ when spin rotation occur. The parameter $\mu$ is a fitting parameter of our model.

In contrast to that of the ground state, the excited state's lifetime is short. In this case, the flipping probabilities strongly depend on the duration of the excited state's evolution time, since it can be shorter than the periodicity of the oscillatory probabilities $p^{\text{ES}} \! \left( \left. \chi_{i} \right| \chi_{f} \right)$, see Fig.~\ref{fig:NV-tev}. For a proper description, we have to include time in our considerations. 

The flipping probabilities, correspond to the excited state's time evolution, thus
\begin{equation} \label{eq:p_es}
p^{\text{ES}} \! \left(  \left. \chi_{\text{Initial}} \right| \chi_{\text{Final}} \right) = \int^{T_{\text{ES}}}_{0} \varrho \! \left( t\right) \left| \left\langle \chi_{\text{Final}} \right| e^{-i \hat{H}^{\text{ES}} t / \hbar} \left| \chi_{\text{Initial}} \right\rangle \right|^{2} dt\text{,}
\end{equation}
where $ \varrho\! \left( t\right)$ is the probability distribution function of the effective length of the excited state's evolution time. $ \varrho\! \left( t\right)$ is assumed to be an exponential distribution
\begin{equation}
\varrho \! \left( t \right) = \frac{1}{\tau_{\text{ES}}^{*}} e^{- t / \tau_{\text{ES}}^{*}}\text{,}
\end{equation}
where $\tau_{\text{ES}}^{*}$ is the characteristic time of the decay. In our model, $\tau_{\text{ES}}^{*}$ is the average effective time of evolution in the excited state, which is considered to be proportional to the excited state's lifetime  $\tau_{\text{ES}}$,
\begin{equation}
\tau_{\text{ES}}^{*} = \nu \tau_{\text{ES}}\text{.}
\end{equation}
Here, $\nu$, which is the last free parameter of our model, is a scaling factor that takes into account electron spin relaxation effects that can change the net evolution time.

Finally, as discussed during the description of the model, the preferential direction $\mathbf{e}$ of the nuclear spin polarization may deviate from the $C_{3v}$ axis of the defect \cite{Fischer2013}. We determine this direction from the ground-state-spin Hamiltonian by finding the direction $\mathbf{e}$, where the nuclear spin state $\left| \uparrow_{\mathbf{e}} \right\rangle$, which is an eigenstate of $\hat{I}_{\mathbf{e}}$ with $+1/2$ eigenvalue, possess the lowest probability to flip into $\left| \downarrow_{\mathbf{e}} \right\rangle$ in the ground state. This means that the precession of state $\left| \uparrow_{\mathbf{e}} \right\rangle$ is minimal and this state is therefore the stationary state of the dynamic nuclear spin polarization cycle. The criterion used is
\begin{equation} \label{eq:finde}
\min_{\mathbf{e}} p^{\text{GS}} \! \left( \left. 0 \uparrow_{\mathbf{e}} \right| 0 \downarrow_{\mathbf{e}} \right)\text{.}
\end{equation}

The calculation of the nuclear spin polarization $P$ for a given system defined by the spin Hamiltonian of Eq.~(\ref{eq:GS_Ham}) can now be carried out by finding the suitable nuclear spin state $\left| \uparrow \right\rangle$ and $\left| \downarrow \right\rangle$ by Eq.~(\ref{eq:finde}), and using it to determine the flipping probabilities of the spin state in accordance with Eqs.~(\ref{eq:p_gs}) and (\ref{eq:p_es}). From these values, the total probability of nuclear spin down-to-up and up-to-down flipping can be obtained by Eq.~(\ref{eq:p_pm}) which provides the polarization of the nuclear spin via Eq.~(\ref{eq:pol}).

The predefined parameters of the model are the diagonal elements and angles of the excited and ground state's hyperfine tensors, $A^{\text{GS}}_{xx}$, $A^{\text{GS}}_{yy}$, $A^{\text{GS}}_{zz}$, $\theta_{\text{GS}}$, $A^{\text{ES}}_{xx}$, $A^{\text{ES}}_{yy}$, $A^{\text{ES}}_{zz}$, and $\theta_{\text{ES}}$, the zero-field-splitting parameter of the ground and excited state's zero-field-splitting tensors, $D_{\text{GS}}$ and $D_{\text{ES}}$, the rate of non-radiative decay $\Gamma$,  and the excited state's lifetime $\tau_{\text{ES}}$. The free parameters used to fit the theoretical curves to the experimental ones are $\mu$, $\nu$, and $\kappa$.

\section{\emph{Ab initio} calculations of the model parameters for the NV center in diamond and the divacancy in 4H and 6H SiC}
\label{sec:abinitio}

Since some key parameters of the DNP model have not been measured, we apply \emph{ab initio} methods to calculate them. In particular, we calculate the full hyperfine tensor of selected proximate $I=1/2$ isotopes in the ground and excited states. In addition, we identify the axial divacancy configurations in 6H-SiC by calculating their $D_\text{GS}$ parameters in order to provide a direct comparison for DNP processes in 6H-SiC. 

We carry out first-principles density-functional theory (DFT) calculations to study NV center in diamond and axial divacancies in 4H and 6H SiC. We apply 512-atom supercell for diamond, and 576-atom and 432-atom supercells for 4H and 6H SiC, respectively. We apply $\Gamma$-point sampling of the Brillouin-zone, which suffices to ensure convergent charge and spin densities.  We utilize the plane wave basis set together with projector augmented wave method as implemented in \textsc{VASP5.3.5} code \cite{VASP, VASP2, PAW, Kresse99}. We apply our in-house code to calculate the GS zero-field splitting from DFT wave functions \cite{Falk2014, Ivady2014}, which has been tested and shown to provide good results with using Perdew-Burke-Ernzerhof (PBE) functional \cite{PBE}. We apply the HSE06 hybrid functional \cite{HSE03, HSE06} to calculate the hyperfine tensors of selected $I=1/2$ nuclei where the spin polarization of the core electrons are taken into account \cite{Szasz2013}. We apply the constrained DFT method to calculate the ES spin density \cite{Gali:PRL2009}.

\subsection{Results on NV center in diamond}

The full hyperfine tensors of $^{15}$N and $^{13}$C coupled to NV centers in diamond have not been experimentally determined. We thus apply \emph{ab initio} calculations to obtain these parameters. The GS-hyperfine tensors of NV center in diamond have been recently characterized in detail  \cite{Szasz2013}. In particular, we analyze the hyperfine tensors in the GS and ES for coupled $^{15}$N nuclei, for which the DNP was thoroughly studied experimentally\cite{Jacques2009}.  We provided the ES hyperfine tensor by PBE functional \cite{Gali2009}, where we showed that the hyperfine constants are anisotropic for $^{15}$N. We list the corresponding HSE06 values in GS and ES in Table~\ref{tab:NVhf}.

\begin{table}[h!]
\caption{\label{tab:NVhf} The calculated hyperfine tensors for the NV center in diamond in the ground (GS) and excited (ES) states. The hyperfine constants ($A_{xx}$, $A_{yy}$, $A_{zz}$) are shown as well as the direction cosine of the largest $A_{zz}$ hyperfine constant represented by angle $\theta$, which is the angle between the direction of $A_{zz}$ and the symmetry axis. The $A_z$ hyperfine constant is the projected hyperfine tensor onto the symmetry axis. The sites are defined by  Smeltzer \emph{et al.} \cite{Smeltzer2011}.
}
\begin{ruledtabular}
\begin{tabular} {lccc|cc }
Nucleus$_{\text{site}}$, state  & $A_{xx}$ (MHz) & $A_{yy}$ (MHz) & $A_{zz}$ (MHz) & $A_{z}$ & $\theta$ ($^{\circ}$)  \\ \hline
  $^{15}$N, GS & 3.9 & 3.9 & 3.4 & 3.8 & 0\\
  $^{15}$N, ES & -38.5 & -38.5 & -58.1 & -46.0 & 0\\ \hline
  $^{13}$C$_{\text{a}}$, GS & 114.0 & 114.1 & 198.4 & 147.6 & 71.7\\
  $^{13}$C$_{\text{a}}$, ES & 44.8 & 45.0 & 117.5 & 77.0 & 69.1 \\ \hline
  $^{13}$C$_{\text{A}}$, GS & 12.7 & 12.8 & 18.5 & 14.9 & 72.0\\
  $^{13}$C$_{\text{A}}$, ES & 9.6 & 9.7 & 15.1 & 11.8 & 73.2\\ \hline
  $^{13}$C$_{\text{B}}$, GS & 11.5 & 11.6 & 17.0 & 13.6 & 68.3 \\
  $^{13}$C$_{\text{B}}$, ES & 9.7 & 9.7 & 15.3 & 11.8 & 64.8\\ \hline
  $^{13}$C$_{\text{C}}$, GS & -10.3 & -10.5 & -8.4 & -10.0 & 28.1\\
  $^{13}$C$_{\text{C}}$, ES & -6.9 & -7.4 & -3.6 & -6.2 & 13.5\\ \hline
  $^{13}$C$_{\text{D}}$, GS & -6.8  & -7.2 & -3.8 & -6.1 & 71.8\\
  $^{13}$C$_{\text{D}}$, ES & -7.4 & -7.8 & -5.3 & -6.9 & 79.1\\ \hline
  $^{13}$C$_{\text{E}}$, GS & 2.9  & 3.0 & 4.8 & 3.6 & 29.3\\
  $^{13}$C$_{\text{E}}$, ES & 0.7 & 1.4 & 2.5 & 1.6 & 23.1\\ \hline
  $^{13}$C$_{\text{F}}$, GS & 4.5  & 4.9 & 2.9 & 4.2 & 55.1\\
  $^{13}$C$_{\text{F}}$, ES & 3.2 & 3.8 & 4.8 & 4.0 & 40.1 \\ \hline
  $^{13}$C$_{\text{G}}$, GS & 2.1  & 2.2 & 3.5 & 2.6 & 75.3\\
  $^{13}$C$_{\text{G}}$, ES & 1.3 & 1.3 & 2.4 & 1.7 & 77.4\\ \hline
  $^{13}$C$_{\text{H}}$, GS & 1.0  & 1.0 & 2.0 & 1.4 & 15.0\\
  $^{13}$C$_{\text{H}}$, ES & 2.0 & 2.0 & 3.5 & 2.6 & 13.9\\
\end{tabular}
\end{ruledtabular}
\end{table}

\subsection{Results on divacancies in 4H and 6H SiC}

We calculate the electronic structure of the axial divacancy defects in 4H- and 6H-SiC in their neutral charge state with $S=1$ electron spin. We have previously determined the $D_\text{GS}$ parameter for $hh$ and $kk$ divacancies in 4H SiC\cite{Falk2014}, and now we report it for $hh$, $k_1k_1$ and $k_2k_2$ divacancies in 6H-SiC. We list the results in Table~\ref{tab:D}. The results imply that QL1, QL2, and QL6 ODMR signals are associated with $k_1k_1$, $hh$, and $k_2k_2$ divacancies, respectively.  
\begin{table}[h!]
\caption{\label{tab:D} A summary of the spin-transition energies for the c-axis-oriented PL6 defect and neutral divacancies in 4H- and 6H-SiC. The parameters are all 20~K parameters, except for the $D_\text{ES}$ of PL6, where the room-temperature value is given. Both $D_\text{GS}$ and $D_\text{ES}$ are positive. Comparing the experimental $D_\text{GS}$ with the calculated $D_\text{GS}$ (calculated at $T$=0~K, using the method in Refs.~\onlinecite{Falk2014, Ivady2014}) allows each spin resonance transition in 6H-SiC to be corresponded with its form of neutral divacancy.}
\begin{ruledtabular}
\begin{tabular}{lccc}
Defect & $D_\text{GS}$ (GHz) & $D_\text{GS}^\text{calc}$ (GHz) & $D_\text{ES}$ (GHz) \\ \hline
4H: $hh$ & 1.336 & 1.358 & 0.84 \\	
4H: $kk$ & 1.305 & 1.320 & 0.78 \\
4H: PL6  & 1.365 &     --    & 0.94 \\
6H: $hh$ & 1.334 & 1.350 & 0.85 \\
6H: $k_1k_1$ & 1.300 & 1.300 & 0.75 \\
6H: $k_2k_2$ & 1.347 & 1.380 & 0.95 \\
\end{tabular} 
\end{ruledtabular}
\end{table}

Proximate $^{29}$Si nuclear spins of divacancies in GS have been detected by means of ODMR \cite{FalkPRL2015}, labeled as Si$_\text{IIa}$ and Si$_\text{IIb}$ by following the labels in Ref.~\onlinecite{Son2006} (see Fig.~\ref{fig:V2}). 
The $hh$ and $k_2k_2$ divacancies show similar GS hyperfine constants, whereas the $k_1k_1$  divacancy exhibits larger values in 6H SiC (see Table~\ref{tab:V2hf}). The $k_1k_1$ in 6H-SiC and $kk$ in 4H-SiC  as well as $hh$ configurations in 6H- and 4H-SiC show similar values. These trends support the assignment of QL1,2,6 based on the calculated $D_\text{GS}$ parameters. 

Finally, we calculated the corresponding hyperfine tensors in the ES (see Table~\ref{tab:V2hf}). Within the accuracy of measurements, the ES shows signatures of $C_{3v}$ symmetry. However, a full characterization of the divacancy's ES is beyond the scope of this study. We approximate the ES by the constraint DFT procedure and fix the $C_{3v}$ symmetry in our calculations to obtain the ES hyperfine tensors.     

\begin{table}[h!]
\caption{\label{tab:V2hf} The calculated hyperfine tensors for nuclear spins that are proximate to divacancies in 6H-SiC in the ground (GS) and exited (ES) state.  The hyperfine constants ($A_{xx}$, $A_{yy}$, $A_{zz}$) are shown, and also the direction cosine $\theta$, which is the angle between the direction of $A_{zz}$ and the symmetry axis. 
The $A_z$ hyperfine constant is the projected hyperfine tensor onto the symmetry axis.
The atom labels are shown in Fig.~\ref{fig:V2}. Since PL6 in 4H-SiC has not yet been identified, we applied the calculated hyperfine tensors in the $k_2k_2$ divacancy configuration in the DNP simulations. In our experience, the difference in the measured or calculated hyperfine constants are within 1~MHz, and a 1~MHz inaccuracy does not alter the simulation results.}
\begin{ruledtabular}
\begin{tabular} {llll|ccc|c|c}
 Nucleus & Site & Conf. & St.  & $A_{xx}$ [MHz] & $A_{yy}$ [MHz] & $A_{zz}$ [MHz]& $\theta$ [$^{\circ}$]  & $A_{z}$ [MHz]    \\ \hline
$^{29}$Si & Si$_{\text{IIb}}$ & 6H: $hh$ & GS & 9.8 & 8.6 & 10.7 & 69.5 & 9.6  \\ 
$^{29}$Si & Si$_{\text{IIb}}$ & 6H: $hh$ & ES & 9.8 & 9.3 & 10.4 & 63.0 & 9.8  \\  
$^{29}$Si & Si$_{\text{IIb}}$ & 6H: $k_1k_1$ & GS & 10.7 & 9.8 & 11.5  & 70.8 & 10.5 \\ 
$^{29}$Si & Si$_{\text{IIb}}$ & 6H: $k_1k_1$ & ES & 10.2 & 9.5 & 10.8  & 60.7 & 10.1 \\ 
$^{29}$Si & Si$_{\text{IIb}}$ & 6H: $k_2k_2$ & GS & 9.9 & 8.7 & 10.8 & 69.4 & 9.7 \\ 
$^{29}$Si & Si$_{\text{IIb}}$ & 6H: $k_2k_2$ & ES & 10.4 & 9.7 & 11.0 & 64.1 & 10.3 \\ 
\end{tabular}
\end{ruledtabular}
\end{table}

\section{Results and discussion on dynamic spin polarization processes}
\label{sec:dnp}

In this section, we present our results and conclusions on the DNP process for the NV center in diamond and for the divacancy in 6H-SiC. First, we start with the study of the NV center in diamond, which is the most thoroughly investigated defect system exhibiting DNP. We show that the new model is capable of reproducing both theoretical and experimental curves. Nevertheless, it provides a new and deeper insight to the features of the DNP mechanism and to the physics of the defect. After this, we apply our model for different configurations of the divacancy in 6H-SiC. Our results reveal the importance of the electron spin coherence time in the excited state in understanding DNP.  

\subsection{NV-center in diamond and a single adjacent $^{15}$N nuclear spin}
\label{sec:NV}

Previous models\cite{Jacques2009} successfully explained experimental results on the dynamic nuclear spin polarization of a $^{15}$N nuclear spin of the NV-center in diamond. In this special case, the hyperfine tensor of the excited and ground states are symmetric (Table~\ref{tab:NVhf}). Therefore, when the magnetic field is well aligned with the axis of the defect the spin Hamiltonian becomes rather simple \cite{Jacques2009,Gali2009}. Here, we reassess these experimental and theoretical results in the framework of our extended method.

 \begin{figure}
\includegraphics[width=0.9\columnwidth]{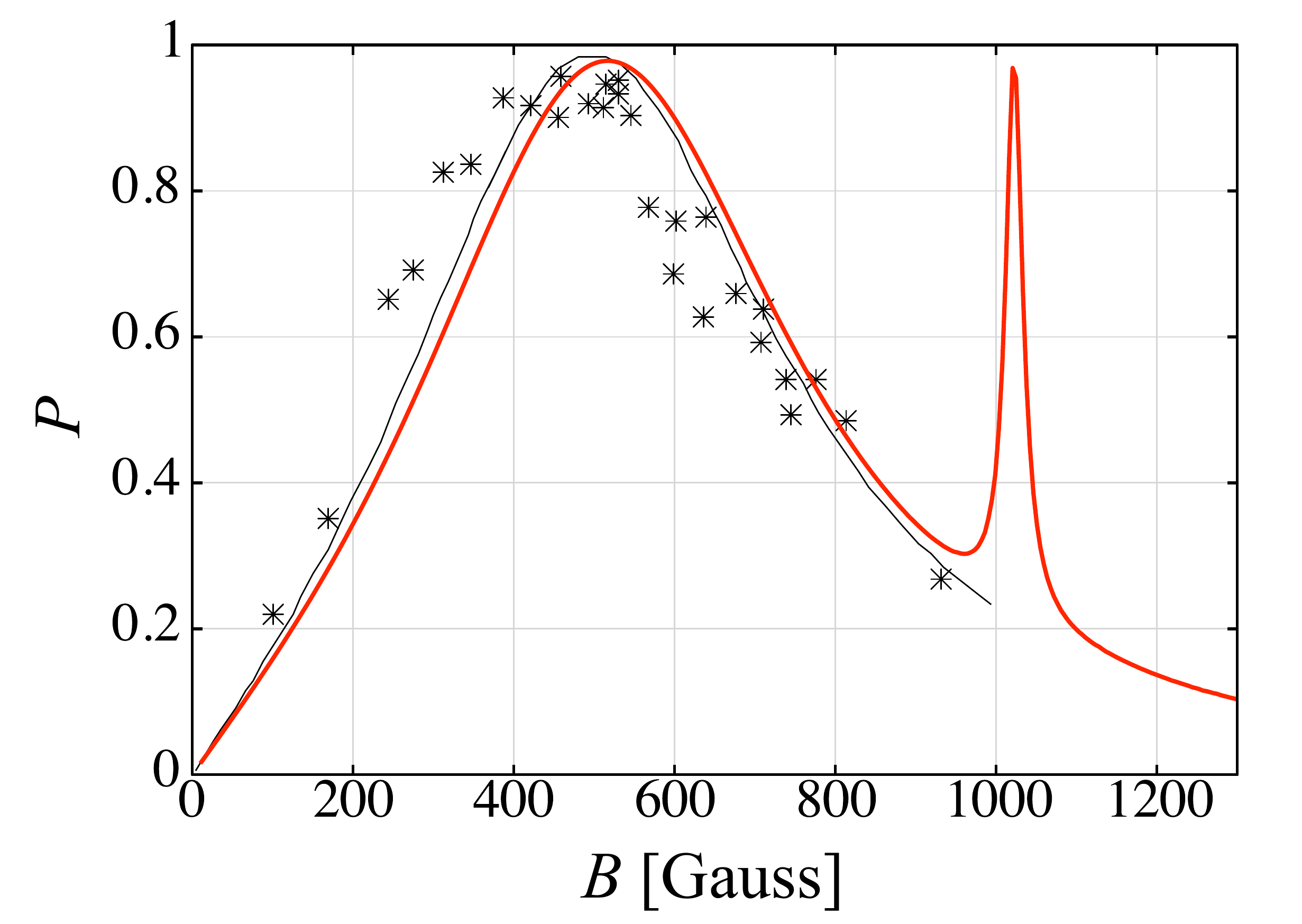}
\caption{\label{fig:NV-P-par} A comparison of calculated and measured dynamic nuclear spin polarization $P$ of a $^{15}$N nucleus of the NV-center in diamond as a function of the external magnetic field $B$. The result of our calculation is depicted with a thick red solid line, while the previous theoretical and experimental results\cite{Jacques2009} are depicted with thin black solid line and black points, respectively. The calculated ESLAC resonance peak, at $B_{\text{ESLAC}} = 516$~G, resembles the reported ones\cite{Jacques2009}. However, in our calculation the ground state processes are also taken into account, producing an additional peak at $B_{\text{GSLAC}} = 1020$~G.}
\end{figure}

To calculate the magnetic field dependence of the nuclear spin polarization, we chose parameters for our model as follows. The parameters of the hyperfine tensor of the ground and excited state are determined by our first principles calculations, Table~\ref{tab:NVhf}. The $D$ parameters for the ground and excited state zero-field-splitting tensors were set to 2.87 and 1.42~GHz, respectively, in accordance with experiment. For the excited state the lifetime and the rate of non-radiative decay, we used the experimental values $\tau_{\text{ES}} =12$~ns and $\Gamma = 0.3$. The $\mu$ parameter in our model determines the features of the GSLAC resonance peak. Given the lack of experimental data for this peak, we set $\mu=0.25$. The shape of the ESLAC polarization curve is determined by two parameters in our model. Parameters $\kappa$ and $\nu$ take into account the effect of the nuclear spin-lattice relaxation and the excited state's electron spin decoherence, respectively. These parameters were fit to the experimental curves $P\! \left( B\right)$ and $P\! \left( \Theta_{B} \right)$. The optimal values are $\nu=0.13$ and $\kappa = 3.15 \times 10^{-4}$.

 \begin{figure}
\includegraphics[width=0.9\columnwidth]{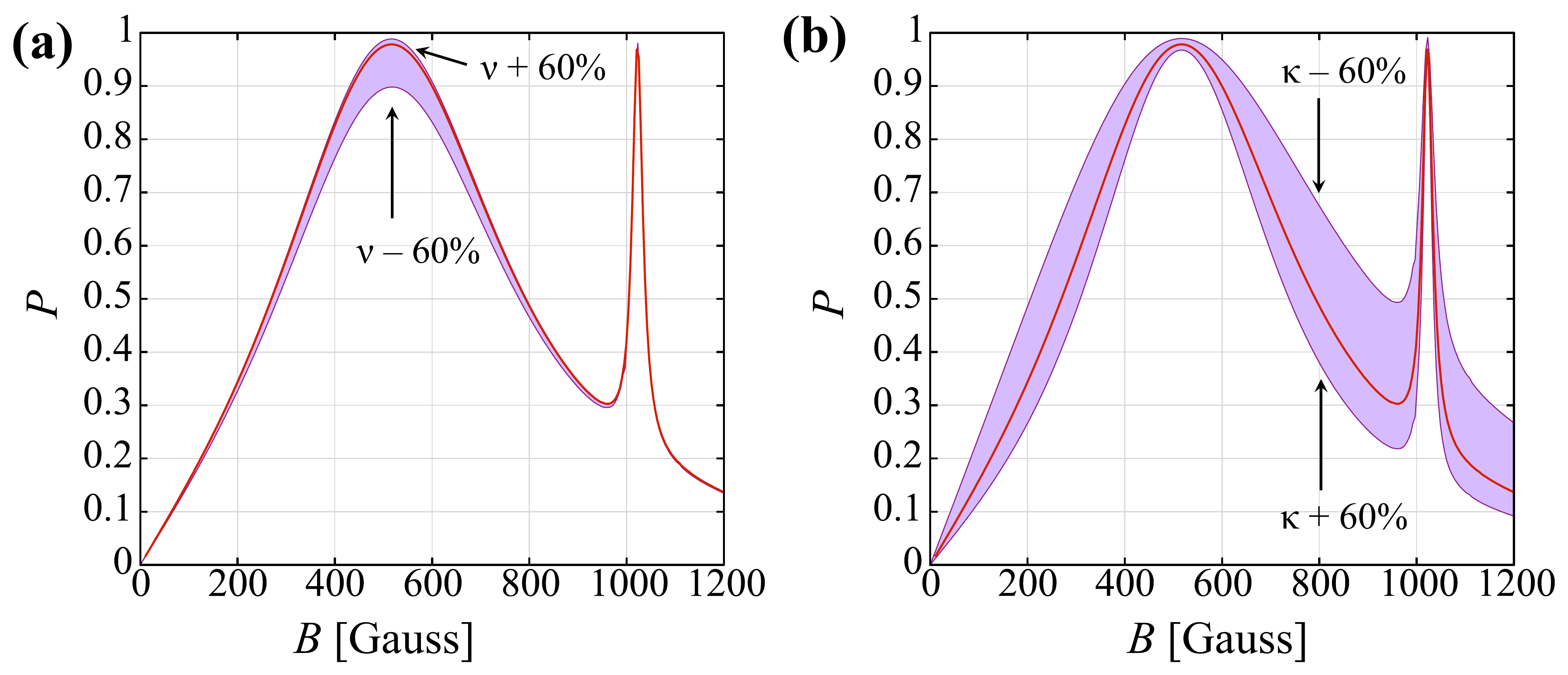}
\caption{\label{fig:NV-fitvar} Parameter dependence of the theoretical nuclear spin polarization curve $P \! \left( B \right)$ for the case NV center in diamond including an $^{15}$N nuclei. (a) and (b) show $\nu$ and $\kappa$ parameter dependence of the theoretical curve, respectively. In both cases, the red thick curve corresponds to the optimal parameter setting, see Fig.~(\ref{fig:NV-P-par}), while the edges of the light purple filled area correspond to $\pm 60 \%$ change of the parameters. Variation of parameter $\mu$ has similar effect on the GSLAC peak as $\nu$ has on the ESLAC peak.}
\end{figure}

In order to indicate the sensitivity of the fitted curve to the variation of the free parameters, we depicted curves of modified $\nu$ and $\kappa$ parameters in Fig.~(\ref{fig:NV-fitvar}).  As can be seen the theoretical curve moderately but non-linearly varies with the changes of the fitting parameters. 

We note that the fitting parameter $\kappa$ has a smaller value in our model than parameter $k^{0}_{\text{eq}}=0.0027$ of the previous model with same definition. The difference is due to the effect of the ES electron spin decoherence. In the previous model, this effect is not included, while in our model it is taken into account explicitly. The electron spin decoherence shortens the effective ES lifetime, thereby reducing the probabilities of nuclear spin flips and the polarization. This finding immediately draws attention to the importance of the electron spin decoherence.   

Our extended model reproduces well the reported magnetic field dependence of the nuclear spin polarization\cite{Jacques2009} at the vicinity of $B_{\text{ESLAC}}$, see Fig.~\ref{fig:NV-P-par}. The position of the maximum polarization is at $B = 516$~Gauss. It is slightly shifted due to the different hyperfine tensors utilized in the two calculations. On the other hand, in our model, the ground state processes are simultaneously described that produces an additional resonance peak at $B_{\text{GSLAC}} = 1020$~Gauss. As the ground state hyperfine interaction is weaker than the excited state interaction, the GSLAC resonance peak is much sharper that at the ESLAC.   

\begin{figure}
\includegraphics[width=0.9\columnwidth]{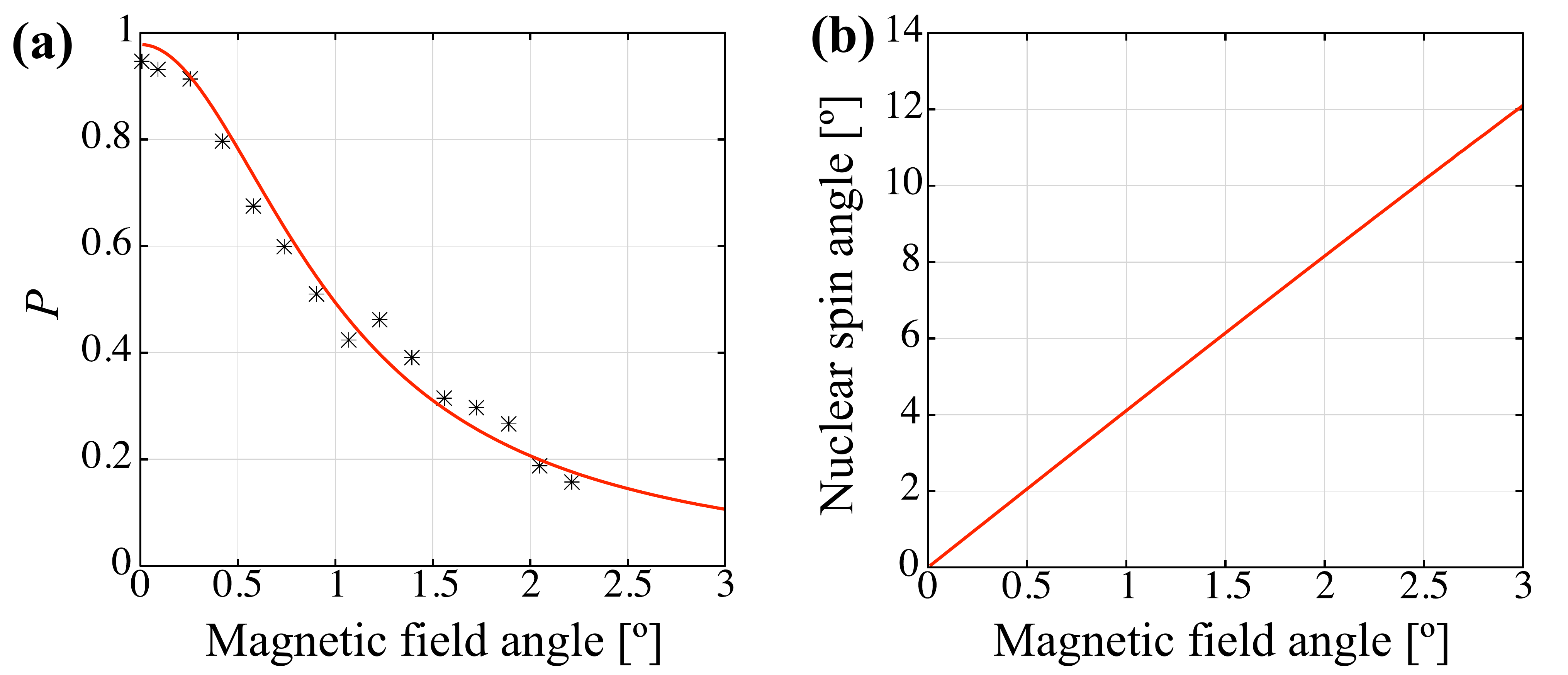}
\caption{ \label{fig:NV-angle} The magnetic field angle dependence of the nuclear spin polarization. (a) shows the calculated (solid line) and measured\cite{Jacques2009} (points) spin polarization of a $^{15}$N nucleus of an NV-center in diamond as a function of the angle of the external magnetic field and the $C_{3v}$ axis of the defect at $B=516$~Gauss. Our calculation accurately reproduces the experimental observations. (b) The angle of the nuclear spin is depicted as function of the angle of the magnetic field. As the misaligned magnetic field reduces the symmetry, the stationary state of the nuclear spin points out of the $C_{3v}$ axis.}
\end{figure}

We also calculated the magnetic field angle dependence of the degree of $P$, see Fig.~\ref{fig:NV-angle} which was not considered by previous models. The theoretical curve agrees well with the result of the experimental measurements\cite{Jacques2009}. The polarization decays very quickly as the angle of the magnetic field increases. Our model makes it possible to investigate these observations thoroughly. First of all, as the symmetry of the hyperfine tensor is reduced by the non-zero angle of the magnetic field and the symmetry axis of the defect in the stationary state, the nuclear spin is not parallel with either the symmetry axis of the defect or the magnetic field. The angle $\theta_{\text{P}}$ of the preferential direction of the polarization linearly depends on the angle $\theta_{\text{B}}$ of the magnetic field, see Fig.~\ref{fig:NV-angle}. The ratio of the two angles is $\theta_{\text{P}} / \theta_{\text{B}} = 4.03$.

\begin{figure}
\includegraphics[width=0.9\columnwidth]{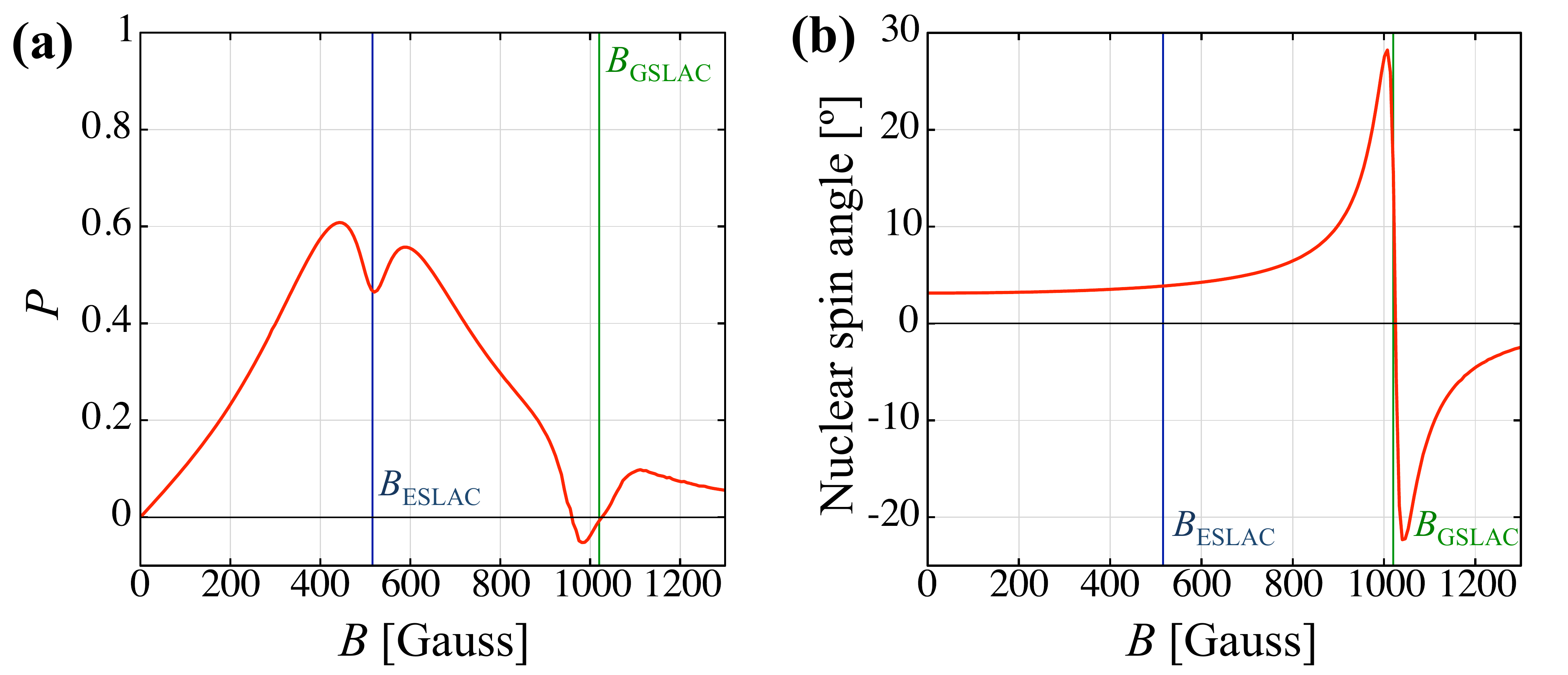}
\caption{\label{fig:NV-P-out} The calculated degree of dynamic nuclear spin polarization of the nucleus $^{15}$N of the NV-center in diamond for the case of misaligned magnetic field. The angle of the magnetic field and the $C_{3v}$ axis of the defect is set to 1$^{\circ}$.  (a) The polarization of the nuclear spin as a function of the strength of the magnetic field. It shows a strong reduction in the polarizability compared with Fig.~\ref{fig:NV-P-par}, due to the misalignment of magnetic field. Interestingly, the ground-state dynamic nuclear spin polarization almost completely disappears. (b) The angle of the nuclear spin and the $C_{3v}$ axis of the defect as a function of the magnetic field.}
\end{figure}

In Fig.~\ref{fig:NV-P-out}, we depicted the magnetic field dependence of the nuclear spin polarization for the case of a misaligned magnetic field. The angle of the magnetic field and the symmetry axis of the defect is set to 1$^{\circ}$. The ESLAC resonance peak is strongly reduced, whereas the GSLAC resonance peak almost disappears.

\begin{figure}
\includegraphics[width=0.9\columnwidth]{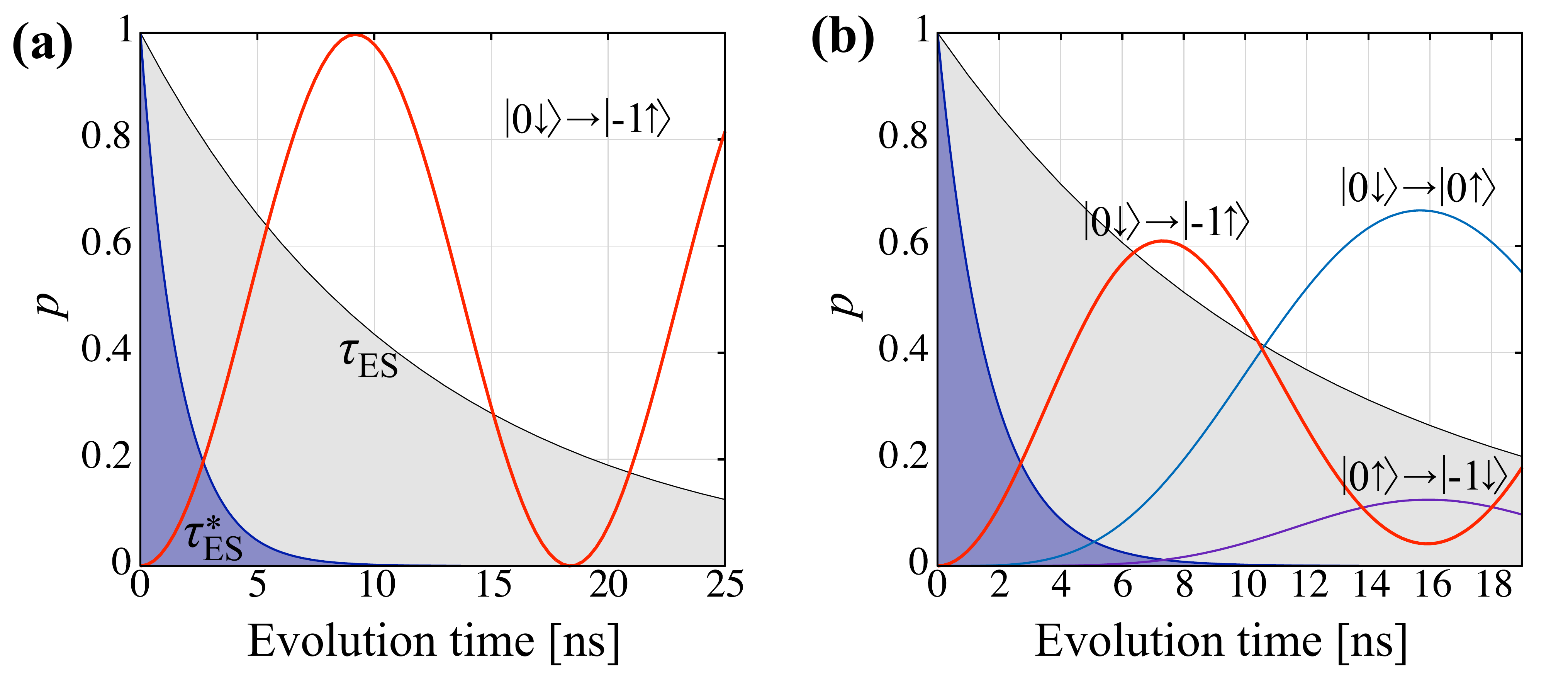}
\caption{\label{fig:NV-tev} The time dependence of spin flipping probabilities in the excited state of the NV center in diamond with a $^{15}$N nucleus. (a) The magnetic field is well aligned with the axis of the defect. In this case, only one spin rotation can occur, $\left| 0 \downarrow\right\rangle \rightarrow \left| -1 \uparrow\right\rangle$, represented with red (thick gray) line. The black and blue lines with light gray and blue filled areas represent the exponential decay of the excited state lifetime and the effective evolution time, respectively (see text for more explanation). The mean evolution time can be seen to be much shorter than the periodicity of the oscillatory probability. (b) The magnetic field is misaligned by 1$^{\circ}$. In this case, different kinds of spin rotations occur that lower the polarizability of the nuclear spin (Not all processes are depicted).}
\end{figure}
 
Next, we investigate the role of different spin rotation processes when the symmetry is reduced. In Fig.~\ref{fig:NV-tev}, we compare the time dependence of the ES spin flipping processes for the case of aligned and misaligned magnetic fields. In the former case, only one spin rotation mechanism can be observed, $\left| 0 \downarrow\right\rangle \rightarrow \left| -1 \uparrow\right\rangle$, in accordance with the previous models \cite{Jacques2009,Gali2009}. By tilting the direction of the magnetic field, the maximal probability of this process decreases while other flipping processes appear simultaneously with high probability (see Fig.~\ref{fig:NV-tev}(b) ). All of these mechanisms reduce the maximal polarizability of the nuclear spin. For instance, $\left| 0 \uparrow\right\rangle \rightarrow \left| -1 \downarrow\right\rangle$ represents a driving force towards nuclear spin polarization in the opposite direction, while $\left| 0 \downarrow\right\rangle \rightarrow \left| 0 \uparrow\right\rangle$ represents the precession of the nuclear spin that reduces the polarization regardless the direction of the nuclear spin. The interplay of these effects is responsible for the reduction of the ES polarizability. 

From Fig.~\ref{fig:NV-tev}, it is clear that the ES lifetime and the length of intact evolution time play an important role in DNP. By fitting the theoretical curve to the experimental data $P\!\left( \Theta_{B} \right)$, we obtain the average length of the intact evolution time, which turns out to be only 14\% of the ES lifetime. This short time suggests substantial electron spin decoherence in the excited state. In such circumstances, the spin rotation processes have a longer timescale than the net evolution time. There are two main consequences of this. First, the nuclear spin flipping probabilities are largely reduced, since the spin rotations are suppressed. Second, there is a preference over fast processes. For example, in the case of misaligned magnetic field, the fastest mechanism, exhibiting rapid oscillations, can flip the spins with the highest probability during the short time evolution. Thus, at the ESLAC $\left| 0 \downarrow\right\rangle \rightarrow \left| -1 \uparrow\right\rangle$ process is still dominant causing non-zero nuclear spin polarization. On the other hand, by elongating the ES evolution time, this preference relaxes and other slower processes can have more pronounced effects (see Fig.~\ref{fig:NV-tev}(b) ), which would greatly change the nuclear spin polarization pattern. This can be the reason for the disappearance of the nuclear spin polarization at the GSLAC (see Fig.~\ref{fig:NV-P-out} ), where the ground state's evolution time is much longer than that of the excited state.

Finally, we emphasize the strong relation between the magnetic field angle dependence of the nuclear spin polarization and the electron spin decoherence in the ES. Shorter coherence time reduces the maximal polarizability but the decay of the polarization with respect to angle $\Theta_{B}$ of the magnetic field is slower. 

\subsection{NV-center in diamond and first neighbor $^{13}$C nuclear spin}
\label{sec:NVC}

In this section, we study the dynamic nuclear spin polarization of a $^{13}$C nuclei in the first neighbor shell, C$_{a}$ site, of the NV center, where both the ESLAC and GSLAC polarization was observed\cite{Jacques2009,HaiJing2013}. In the latter case, particular features appear due to the non-symmetric hyperfine interaction, which show a potential for new applications in the area of sensitivity-enhanced nuclear magnetic resonance and spintronics \cite{HaiJing2013}.

As a N atom with non-zero nuclear spin is an inherent part of the NV center, the inclusion of an adjacent $^{13}$C nucleus necessarily results in a three-spin system. Due to the small gyromagnetic ratio of the nuclei, the direct nuclear spin-nuclear spin interaction is negligible.  Substantial interaction is, however, mediated by the electron. In this case, the hyperfine interaction with the nitrogen atom appears as an additional magnetic field dependent electron spin relaxation effect for the two-spin system of the nucleus  $^{13}$C and the electron. On the other hand, in the stationary state of the system, the nitrogen nuclear spin is highly polarized, while the populated spin state of the nitrogen nucleus is only weakly coupled to the electron spin. Therefore, we neglect the effect of the nitrogen nuclear spin on the polarizability of the  $^{13}$C nucleus. We investigate the system as a two-spin system within our model. Description of three-spin DNP systems can be found in Ref.~\onlinecite{Nizovtsev2014}.

 \begin{figure}
\includegraphics[width=0.9\columnwidth]{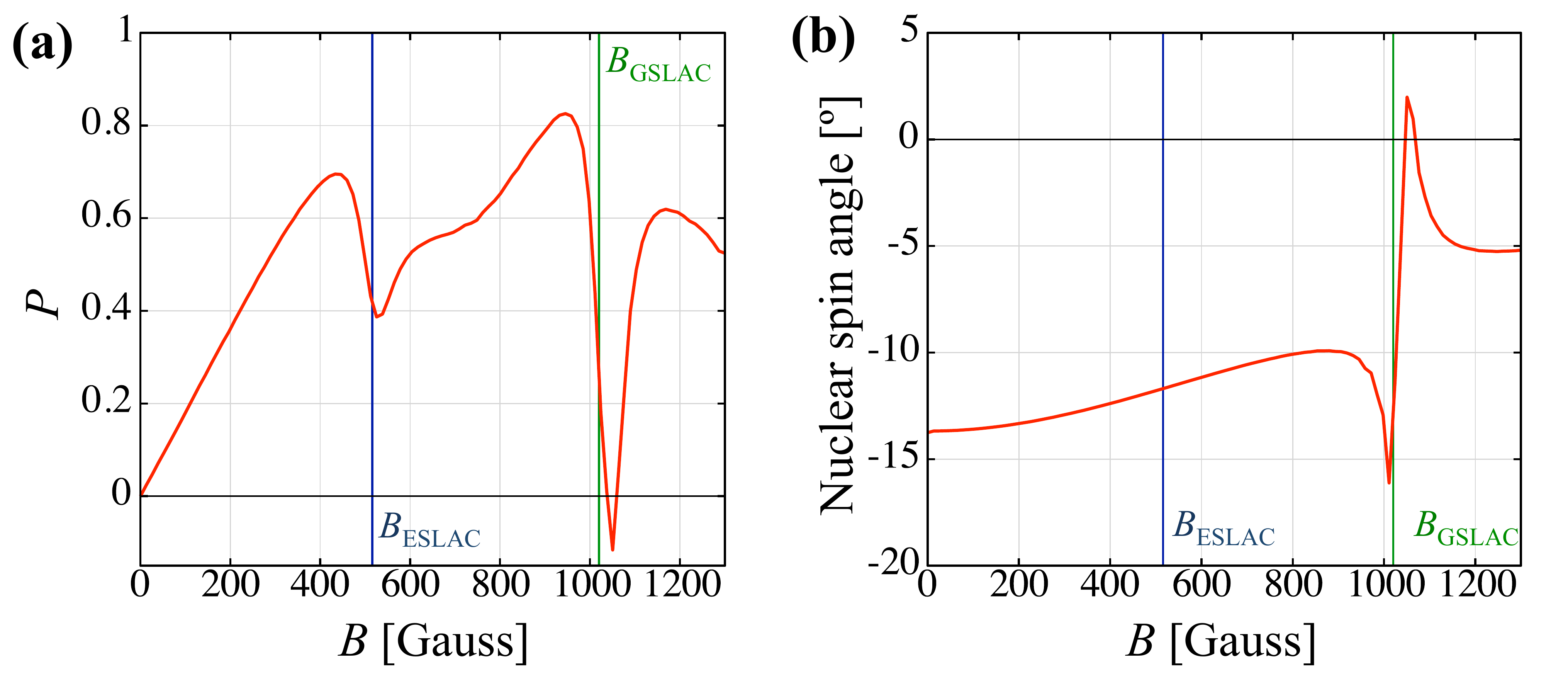}
\caption{\label{fig:NV-Ca} Calculated magnetic field dependence of the dynamic nuclear spin polarization of a single nucleus $^{13}$C in the first neighbor shell of the NV-center in diamond.  (a) The polarization of the nuclear spin as a function of the strength of the magnetic field.  (b) The angle of the nuclear spin and the $C_{3v}$ axis of the defect as a function of the magnetic field.}
\end{figure}

In the calculation, we used the hyperfine parameters presented in Table~\ref{tab:NVhf}. Other parameters are the same as in the previous section. The reason for keeping the parameters is that the measurements on the ESLAC polarizability of $^{14}$N and $^{13}$C nuclei were carried out on same sample\cite{Jacques2009}.

As can be seen in Fig.~\ref{fig:NV-Ca}, the predicted maximal nuclear spin polarization of 70\% at ESLAC is slightly below the experimentally observed 90\% \cite{Jacques2009}.  Since the measurements were carried out on single centers, the local environment is not necessarily the same, which may be the reason for the differences. By assuming weaker spin dephasing effects, higher polarization can be achieved in the calculations. 

The oscillatory features of the calculated nuclear spin polarization of the $^{13}$C nucleus at GSLAC (see Fig.~\ref{fig:NV-Ca}(a) ) reproduces well the experimental observations by Wang \emph{et al}.\ \cite{HaiJing2013}. In our case the nuclear spin polarization reaches 83\% at 950~Gauss before it rapidly drops at GSLAC. Right after the GSLAC, the polarization reverses and reaches -11\% at 1050~Gauss. It then increases again and reaches 62\% at 1170~Gauss. This feature has been thoroughly investigated and understood by Wang \emph{et al}.\ \cite{HaiJing2013}. In short, due to non-symmetric hyperfine interactions, unusual combinations of ladder operators appear in the spin Hamiltonian (see Appendix). From Fig.~\ref{fig:cycle}(a) and (b) it is clear that the reverse of the polarization is due to the dominance of process $\left| 0 \uparrow\right\rangle \rightarrow \left| -1 \downarrow\right\rangle$ over the most common process $\left| 0 \downarrow \right\rangle \rightarrow \left| -1 \uparrow\right\rangle$ in the ground state. This dominant process occurring only at well-defined external magnetic fields is governed by the $\hat{S}_{-} \hat{I}_{-}$ operator combination that resonantly mixes state $\left| 0 \uparrow\right\rangle $ and $ \left| -1 \downarrow\right\rangle$ at a certain magnetic field slightly higher than $B_{\text{GSLAC}}$. The resonance appears as a dip due to the inverse polarization.  Since the hyperfine constant correspond to term $\hat{S}_{-} \hat{I}_{-}$ is smaller than the one correspond to the regular term $\hat{S}_{-} \hat{I}_{+}$, the dip is sharper than the resonance peak due to the term $\hat{S}_{-} \hat{I}_{+}$. In summary, GSLAC polarization curve built up from a wide resonance peak, due to term $\hat{S}_{-} \hat{I}_{+}$, and a slightly shifted sharper dip, due to term $\hat{S}_{-} \hat{I}_{-}$. 

By looking at the polarization near the ESLAC (Fig.~\ref{fig:NV-Ca}), one may recognize similar oscillation of the polarization. However, here the pattern is not as developed as for the GSLAC resonance. From the discussion presented in the previous section, it is clear that the short ES lifetime and the fast electron spin dephasing effect cause the difference. The hyperfine constant of term $\hat{S}_{-} \hat{I}_{+}$ is larger than that of $\hat{S}_{-} \hat{I}_{-}$ \cite{HaiJing2013}, which implies slower spin rotation processes for the latter term. As the short evolution time reduces the flipping capability of slow processes, the dip corresponding to the slower $\hat{S}_{-} \hat{I}_{-}$ process is effectively suppressed. By manually increasing the ES lifetime, we can see a more developed dip.

Finally, the magnetic field dependence of the polarization direction of the $^{13}$C nucleus exhibits similar pattern at the GSLAC (see Fig.~\ref{fig:NV-Ca} (b) ) as $^{15}$N for the case of misaligned magnetic field. One can argue that the two types of symmetry reduction of the spin Hamiltonian result in qualitatively the same phenomena. This realization may provide a way to engineer the magnetic field dependence of the polarization of non-symmetrically placed nuclei.

\subsection{NV-center in diamond and remote $^{13}$C nuclear spins}
\label{sec:NVCC} 

The study of the polarizability of remote $^{13}$C nuclear spins provides an opportunity to have a deep insight into the parameter dependence of the maximal achievable nuclear spin polarization in DNP processes, which is thoroughly examined in the first subsection. Our conclusions are tested on the available experimental results in the second subsection.

\subsubsection{Parameter dependence of nuclear spin polarization}

From the applications point of view, it is important to understand the degree to which the spin polarization of the electrons can be transferred to the nuclei by DNP and what the limiting factors are. Remote nuclear spins couple weakly to the electron spin of the defect, and thus their polarizability is assumed to strongly depend on the details of the hyperfine tensor. In this section, we aim to understand the role of different hyperfine parameters in the determination of the maximal achievable nuclear spin polarization of more remote nuclei that couples relatively weakly to the electron spin of an NV center. Previous publications have already addressed this issue to an extent \cite{Smeltzer2009,Gali2009,Smeltzer2011,Dreau2012}. However, a satisfactory conclusion has not been achieved so far. The most commonly mentioned reasons for the varying polarizability of remote $^{13}$C nuclei are the varying angle of the principal axis of the ES hyperfine tensor and the symmetry axis of the defect as well as the varying angle of the principal axis of the ES and GS hyperfine tensor.  On the other hand, as we will show, the strength of the coupling may have the most important role.

In this section, we investigate a set of $^{13}$C ($I=1/2$) nuclear spins interacting with the electron spin of the NV-center in diamond at ESLAC condition. We examined the achievable nuclear spin polarization as a function of the strength of the coupling, the anisotropy of the hyperfine interaction, the direction cosines of the ES hyperfine tensor, and the difference of the direction cosines of the ES and GS hyperfine tensors. In the calculations, we arbitrarily vary  $A_{xx}^{\text{ES}} = A_{yy}^{\text{ES}}=A_{\parallel}^{\text{ES}}$, and $A_{zz}^{\text{ES}} = A_{\perp}^{\text{ES}}$ in the range of 1--24 MHz and $\theta^{\text{ES}}$ and $\theta^{\text{GS}}$ in the range of 0--90$^{\circ}$ to map the parameter space, while we kept fixed other parameters of the model. The free parameters were initialized as $\kappa = 0.0015$, $\nu = 0.5$, and $\mu =1$.

First, we investigate the effect of GS processes on the ESLAC polarizability of weakly coupled nuclear spins. The GS-spin Hamiltonian plays an important role, since the system spends most of the time in GS. In our model, this Hamiltonian determines the direction of nuclear spin polarization. This suggests that the details of the GS hyperfine interaction can affect the polarizability. Our calculations, however, show that the nuclear spin polarization direction is parallel with the axis of the defect, i.e.\ the nuclear spin polarized parallel to the electron spin, regardless the details of the GS hyperfine tensor. Thus, we can say that the difference of the principal axis of the GS and ES hyperfine tensor, which is thought to play an important role, as well as other parameters of the GS hyperfine tensor, do not affect the ESLAC polarizability of the spin of remote nuclei. One may prove this observation based on simple analytical arguments. Consider the usual basis $\left| 0 \uparrow  \right\rangle$, $\left| 0 \downarrow  \right\rangle$, $\left| -1 \uparrow  \right\rangle$, and $\left| -1 \downarrow  \right\rangle$, where the quantization axis of the electron and nuclear spins are parallel to the symmetry axis of the defect. Far from the GSLAC, the large zero field splitting of the electron spin states suppresses the mixing of the states, while at the GSLAC the mixing is more enhanced due to the GS hyperfine coupling. On the other hand, for the case of weak GS interactions, the mixing and other effects of the interaction are simply negligible at the ESLAC, since $A \approx \mathcal{O}(10 \text{ MHz}) \ll 1400 \text{ MHz} \approx D_{\text{GS}}-D_{\text{ES}}$. Thus, the state $\left| 0 \uparrow  \right\rangle$ and $\left| 0 \downarrow  \right\rangle$ are eigenstates of the GS spin Hamiltonian within good approximation. Their long GS lifetime makes them good target states for DNP polarization. 

\begin{figure}[h!]
\includegraphics[width=0.9\columnwidth]{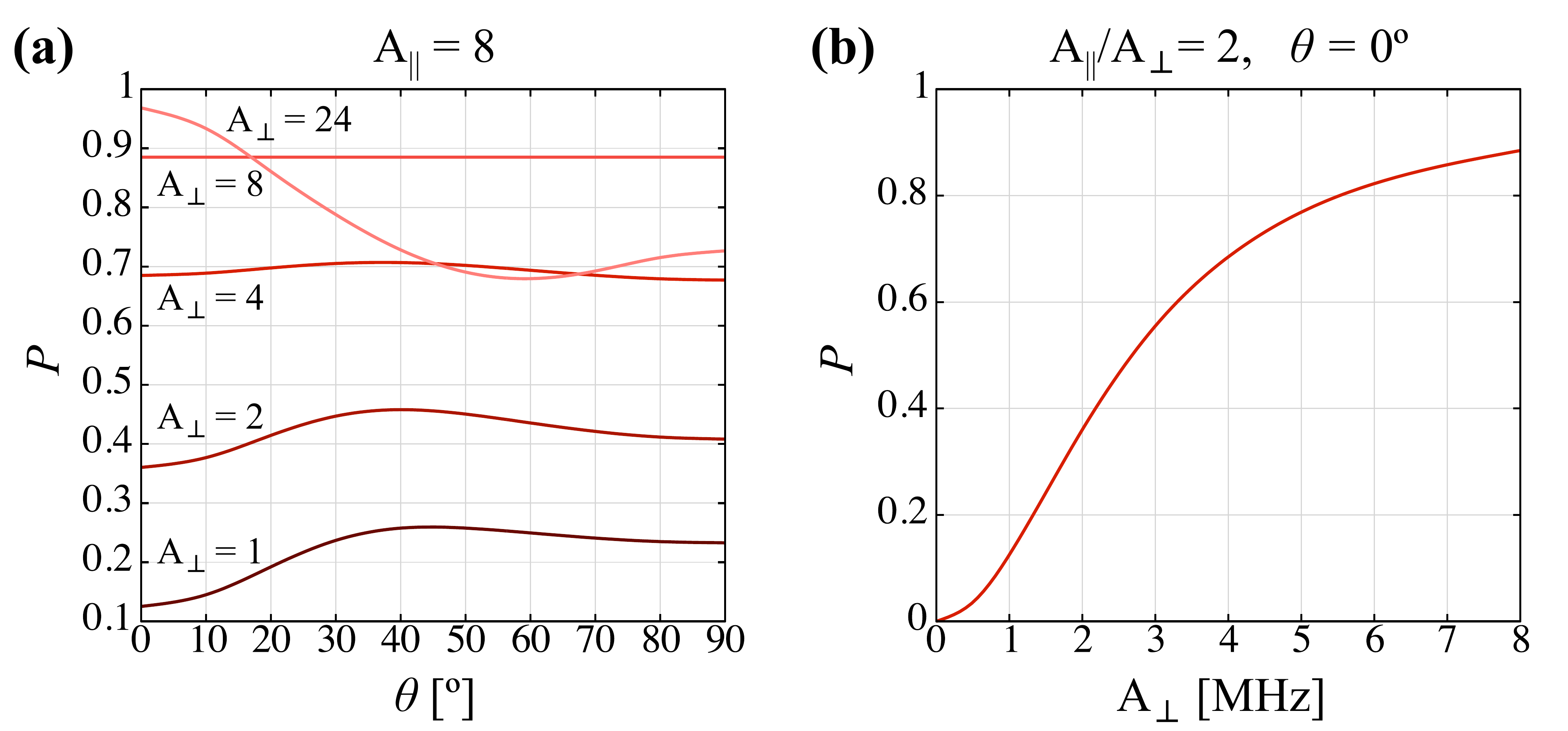}
\caption{ \label{fig:NV-Pardep} Parameter dependence of the dynamic nuclear spin polarization of $^{13}$C nuclei adjacent to a NV center in diamond. (a) The polarization is plotted against the angle of C$_{3v}$ axis and the principal axis of the ES hyperfine tensor for different values of coupling strength $A_{\perp}$.  (b) The polarization is plotted against the coupling strength $A_{\perp}$. The hyperfine parameters are given in MHz unit and refer to ES interactions.}
\end{figure}

To investigate the dependence of the polarizability on the details of the ES hyperfine tensor, we depict the maximal polarization of $^{13}$C nuclear spin as a function of the principal axis of the hyperfine tensor for different coupling strength $A_{\perp}$ in Fig.~\ref{fig:NV-Pardep}~(a). As one can observe, the angular dependence of the polarization is not substantial in most cases. However, it depends on the anisotropy of the hyperfine interaction. For example, when the interaction is isotropic, i.e. $A_{\perp} = A_{\parallel}$, no angular dependence is clearly observed. However, when the interaction is highly anisotropic more pronounced angular dependence can be observed,  see the case of $A_{\perp} / A_{\parallel} = 1/8$ or $A_{\perp} / A_{\parallel} = 3$. To understand the observed features, we need to take into account three effects: First, the value of the off-diagonal elements, which are responsible for the spin flipping processes, are functions of angle $\theta$ as 
\begin{equation} \label{eq:A1}
\left\langle -1 \uparrow \left| \hat{H}_\text{hyp} \right| 0 \downarrow \right\rangle = \frac{\sqrt{2}}{4} \left( A_{\perp} \left( 1 +  \cos^2\left( \theta \right) \right) + A_{\parallel} \sin^2 \left( \theta \right) \right) \text{,}
\end{equation}
For instance, the mixing term increases with angle $\theta$ as $A_{\parallel} > A_{\perp}$, which results in increasing polarizability for small $\theta$, see the case of $A_{\perp} = 1$~MHz and 2~MHz in Fig.~\ref{fig:NV-Pardep} (a). The opposite effect can be observed for $A_{\parallel} < A_{\perp}$. This effect increases with the anisotropy of the hyperfine tensor. Second,  as the angle $\theta$ increases, the symmetry of the spin Hamiltonian becomes more and more reduced, thereby strengthening the nuclear spin depolarization processes (see Fig.~\ref{fig:NV-tev} (b) ). Third, the effect of these processes depends on the net evolution limited by the lifetime of the excited state and the electron spin coherence time in the excited state. The interplay of the above described two basic mechanisms determines the hyperfine angle dependence of the polarizability.

Next, we investigate DNP as a function of the strength of the hyperfine coupling. We depict the maximal polarization against parameter $A_{\perp}$ in Fig.~\ref{fig:NV-Pardep}(b) with fixed $A_{\parallel}/A_{\perp}$ ratio and angle $\theta$. As one can see, the polarization quickly drops as the hyperfine interaction decreases below 10~MHz. The reason of the reduction of the polarizability is the finite evolution time in the ES of the dynamical cycle. As the strength of the coupling decreases, the rate of the spin flipping process decreases as well. Due to the finite lifetime, the time is not sufficient to flip the spins. On the other hand, the character of the curve depicted on Fig.~\ref{fig:NV-Pardep}(b) strongly depends on the length of the effective evolution time and thus on the strength of the electron spin dephasing effects. We can say that the finite lifetime and the short electron spin coherence time of the ES limits DNP of remote nuclear spins. Shorter evolution times causes the drop of the polarizability at larger hyperfine couplings. Here, we mention that the enrichment of the nuclear spin concentration effectively decreases the electron spin coherence time, thus suppressing DNP processes of remote and nearby nuclear spins. This mechanism can have important consequences on the NMR applications of these DNP processes, and may suggest the importance of implementing dynamical decoupling protocols designed to enhance DNP.

To summarize our findings on the hyperfine parameter dependence of the DNP process we can say that, at a given external condition, the polarizability of remote $^{13}$C nuclear spins is determined by their $A_{\perp}$ parameter in the first order. A diverse secondary contribution is governed by the relationship of the three parameters $A_{\perp}$, $A_{\parallel}$, and $\theta$. The details of the GS hyperfine interaction do not affect the polarizability. And finally, the electron spin coherence time limits how far the electron spin polarization can be transferred by ESLAC DNP mechanism.
 
 \subsubsection{Comparison of theory and experiment for the case of remote $^{13}$C nuclei}

\begin{figure}[h!]
\includegraphics[width=0.9\columnwidth]{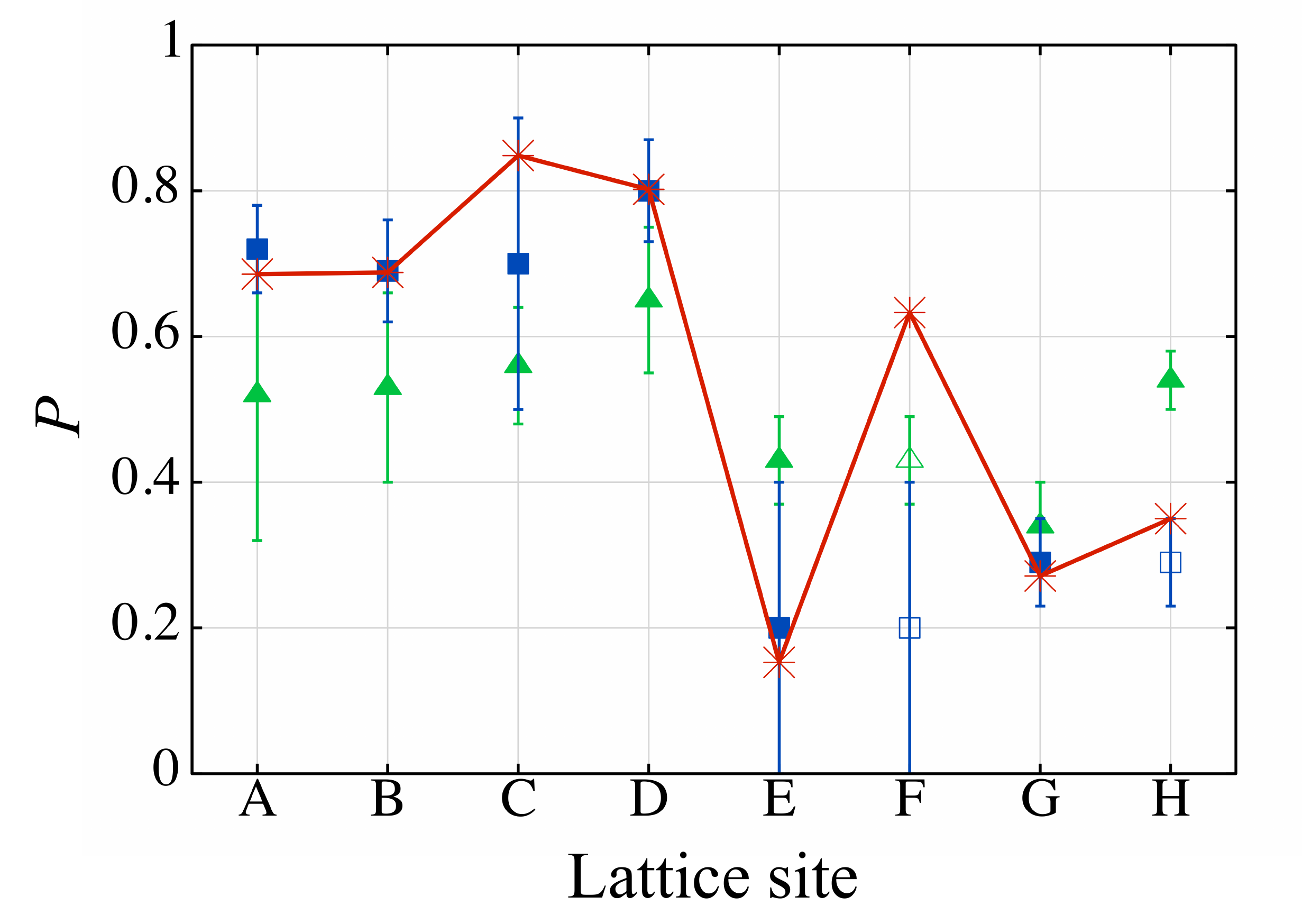}
\caption{ \label{fig:NV-13C} The degree of dynamic nuclear spin polarization of $^{13}$C nuclei at various sites around a NV center in diamond. The calculated polarizations (the red line with stars) are compared with the available experimental results. Blue squares and green triangles show the experimental results by Smeltzer \emph{et al.} \cite{Smeltzer2011} and Dr\'eau \emph{et al.} \cite{Dreau2012}, respectively. The sign of the experimental polarization is changed for sites C and D, see text for more explanation. Sites E and F and sites G and H could not be separated in the experiment due to their similar GS hyperfine parameters. Thus, the measured polarization can be the mixture of the polarization of the two sites. As sites F and H with three-fold degeneracy are more scarce than the corresponding E and G sites with six-fold degeneracy, F and H data points are represented with empty symbols. Because only a few centers with $^{13}$C nuclei on a specific site were investigated, it is possible that only the most probable sites, E and G, were measured.}
\end{figure}

In this subsection, we compare the results of our calculations with the available experimental measurements for the case of distant $^{13}$C nuclei around an NV center in diamond (see Fig.~\ref{fig:NV-13C}). In our calculations, we used the calculated hyperfine coupling parameters listed in Table~\ref{tab:NVhf}, while the zero-field-splitting parameters, the ES lifetime, and the non-radiative decay rate are the same as in the previous sections. The free parameters were specified as $\kappa = 0.003$, $\mu=0.9$, and $\nu=1$. In the calculations the magnetic field was set to $B=510$~Gauss as given in the experiments.

The calculated polarization is in good agreement with the experimental measurements by Smeltzer \emph{et al.} \cite{Smeltzer2011}. However, the agreement is worse for the ones obtained by Dr\'eau \emph{et al.} \cite{Dreau2012} (see Fig.~\ref{fig:NV-13C}). One should keep in mind that there are many factors that may limit the direct comparison between theory and experiments. Beside the experimental uncertainties, the calculated ES hyperfine parameters have systematic uncertainties due to the approximate theory and the difficulties of the description the ES-electron configuration. 

By looking at the theoretical result in Fig.~\ref{fig:NV-13C} as well as at the hyperfine parameters in Table~\ref{tab:NVhf}, the basic roles for the polarizability, defined in the previous subsection, can be recognized. However, there is an additional factor one should take into account. Different $^{13}$C sites exhibit different hyperfine coupling to the electron spin of the NV center. \emph{By applying the same magnetic field for all the different sites the resonance condition is not satisfied equally for all the nuclei spins.} As the hyperfine coupling is relatively weak and the resonance peaks are sharp, the polarization pattern among the different sites can sharply depend on the value of the external magnetic field. This may be a reason for the substantial deviation between the two available experimental measurements. In our calculations, we found that the variation of $\Delta B = \pm5$~Gauss can result in 5--25\% variation of the observed polarization among the different sites. Our previously defined rules apply to maximally achievable polarization and thus to perfect resonance conditions. The imperfect resonance condition overshadows the trends dictated by our findings.

Our calculations show positive polarization for both positive and negative hyperfine parameters. However, in previous manuscripts, negative hyperfine parameters were associated with negative polarization \cite{Smeltzer2011,Dreau2012}. We think that the discrepancy comes from a misinterpretation of the sign of the polarization in the experiments. For positive hyperfine parameters, the population of the lower energy eigenstate represents positive nuclear spin polarization. On the other hand, for negative hyperfine parameters, positive polarization corresponds to the population of the higher energy eigenstates. \emph{In the measurements, the population of the higher energy eigenstate is interpreted as negative polarization. However, it is actually due to the change of the sign of the hyperfine parameter.} Thus, in Fig.~\ref{fig:NV-13C}, we depict the experimental polarization of C and D sites with positive sign, in agreement with our simulations.
 
\subsection{Divacancy in 6H-SiC and adjacent $^{29}$Si nuclear spin}

Recently, dynamic nuclear spin polarization of the $hh$ and $k_{1}k_{1}$ configurations of the divacancy in 6H-SiC and the related PL6 center in 4H-SiC were observed \cite{FalkPRL2015}. In this section, we use our model to reproduce the measured magnetic field dependence of the nuclear spin polarization at the vicinity of the ESLAC and to use the fitting parameters to gain important insight to the DNP processes. Here, we restrict ourselves to study of ESLAC DNP of the different divacancy configurations. 

\begin{figure}[h!]
\includegraphics[width=0.9\columnwidth]{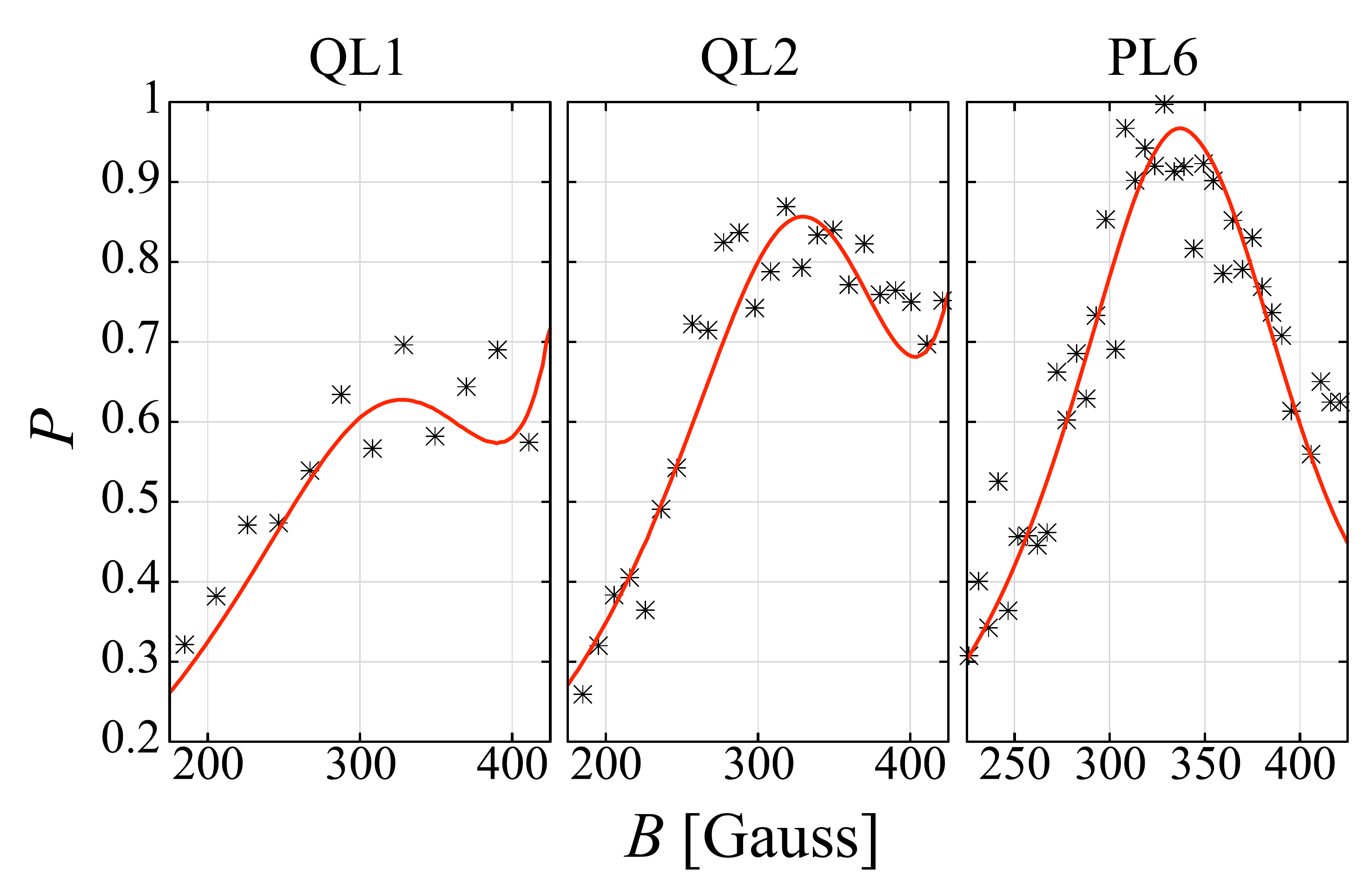}
\caption{ \label{fig:fit-hh} The adjusted theoretical curves (solid lines) and the experimental measurements\cite{FalkPRL2015} (points) on the dynamic nuclear spin polarization of different divacancy configurations and related defects.  The panels show the QL1, QL2 and PL6 photoluminescence centers (ordered from left to right). The first two centers are assigned to the $k_1k_1$ and $hh$ configurations of divacancy in 6H-SiC (see text for further information). The unidentified PL6 center is modeled by hyperfine parameters of $k_2k_2$ configuration in the theoretical simulations.}
\end{figure}

Considering the experimental results\cite{FalkPRL2015} and the calculated hyperfine parameters of the different configurations, see Table~\ref{tab:V2hf}, an interesting observation can be made. Although, the hyperfine parameters of these configurations differ only a few MHz, the measured nuclear spin polarization vary between 60\% and 99\% at ESLAC, see Fig.~\ref{fig:fit-hh}. As we will show in this section, such large variation of the polarizability can only be explained by the different ES electron spin coherence times of the different configurations. This observation further emphasizes the role of these effects in the DNP process. 

\begin{table}[h!]
\caption{\label{tab:D_fit} The utilized excited state zero-field-splitting parameter $D_{\text{ES}}$ as well as the measured \cite{Falk2014} rate of non-radiative decay $\Gamma$ and excited state lifetime $\tau_{\text{ES}}$ are listed for $hh$ and $k_1k_1$ configuration of divacancy in 6H-SiC and the related PL6 defect in 4H-SiC.}
\begin{ruledtabular}
\begin{tabular}{cccc}
Configuration, SiC polytype & $D_\text{ES}$ [MHz]  & $\Gamma$ & $\tau_{\text{ES}} $ [ns] \\ \hline
$hh$, 6H        &  930 & 0.15 & 15 \\
$k_1k_1$, 6H &  915 & 0.16 & 15\\
PL6, 4H          &  955 & 0.33 & 14\\ 	
\end{tabular} 
\end{ruledtabular}
\end{table}

The hyperfine parameters, used in the DNP calculations, were determined by our \emph{ab initio} calculations, see Section~\ref{sec:abinitio}. Recent experiments revealed sensitive temperature dependence of the excited state zero-field-splitting of the divacancy\cite{FalkPRL2015}. Since the DNP measurements were carried out at various temperatures\cite{FalkPRL2015}, we also adjusted $D_{\text{ES}}$ in our DNP simulations. The $E_{\text{ES}}$ parameter is set to zero in all cases. The $D$ parameter together with the measured rate of non-radiative decay $\Gamma$ and ES lifetime $\tau_{\text{ES}}$ are listed in Table~\ref{tab:D_fit}. The fitting parameters correspond to the different type of spin relaxation processes are collected in Table~\ref{tab:Free_fit}.

\begin{table}[h!]
\caption{\label{tab:Free_fit} Fitting parameters $\kappa$ and $\nu$ for $hh$ and $k_1k_1$ configuration of divacancies in 6H-SiC and the related PL6 defect in 4H-SiC. The latter parameter determines the effective evolution time $\tau_{\text{ES}}^{*}$ which is then compared with the experimentally measured \cite{FalkPRL2015} coherence time $T_{2}^{*}$. In addition, the ambient measurement temperature $T$ is given.}
\begin{ruledtabular}
\begin{tabular}{c|c|ccc|c}
Configuration, SiC polytype & $\kappa$  & $\nu$ & $\tau_{\text{ES}}^{*} =\nu\tau_{\text{ES}}$ [ns] & $T_{2}^{*} $ [ns] & $T$ [$^{\circ}$K]\\ \hline
$hh$, 6H        & $8.25\times 10^{-5}$  & 0.092 & 1.38 & 4.0 & 100\\
$k_1k_1$, 6H & $8.25\times 10^{-5}$  & 0.045 & 0.68 & 1.3 & 100\\
PL6, 4H          & $3.14\times 10^{-4}$   & 0.362 & 5.1 & 5.0 & 300 \\ 	
\end{tabular} 
\end{ruledtabular}
\end{table}

The adjusted theoretical curves and the experimental measurements are depicted together in Fig.~\ref{fig:fit-hh} for $hh$ and $k_1k_1$ configuration of divacancy in 6H-SiC and PL6 center in 4H-SiC. Good agreement is achieved for all the cases. We would like to emphasize that the theoretical curves are adjusted by the variation of parameters that are related to spin decoherence and spin-lattice relaxation processes. Through the effect of the variation of these parameters, the effect of different spin relaxation processes may be understood. The $\kappa$ parameter, corresponding to the rate of spontaneous nuclear spin flipping and thus to the nuclear spin-relaxation time, primarily determines the decay of the polarization curve away from the ESLAC resonance. Second, it lowers the maximal achievable polarization at resonance. The $\nu$ parameter corresponds to the effect of electron decoherence in the ES. It mainly determines the height of the resonance peak, thus the maximal polarization at ESLAC. 

Spin relaxation processes are taken into account rather approximately in our model, which might describe either the intrinsic properties of the defects or the different environment of the defects. Considering the fitting parameters in Table~\ref{tab:Free_fit}, one can see that $\kappa$ has larger value for PL6 in 4H-SiC than for $hh$ and $k_1k_1$ in 6H-SiC, suggesting faster nuclear spin-lattice relaxation for PL6. The reason can be threefold: (i) the measurements on the PL6 center were carried out at room temperature while in the other cases at 100 K; (ii) the 4H sample is differently processed than 6H sample with possibly creating different number of disturbing spins the crystals; (iii) PL6 is preferentially located relatively close to the surface \cite{Falk2013} where the concentration of non-coherent electron spins may be larger. In Table~\ref{tab:Free_fit}, one also finds the values of the fitting parameter $\nu$, together with the estimated effective evolution time $\tau_{\text{ES}}^{*}$ with the trend of $k_1k_1 < hh < \text{PL6}$. This trend is good agreement with the one that can be observed from the measured electron dephasing time $T_{2}^{*}$, see Table~\ref{tab:Free_fit}, governing the length of the effective evolution time in the excited state $\tau_{\text{ES}}^{*}$ in our model. This result indicates that the achievable high (low) ESLAC polarizability for PL6  ($k_1k_1$) is due to the longer (shorter) electron coherence time in the ES.

\section{Summary}
\label{sec:conclusion}

We developed a model that provides a detailed description of the dynamic nuclear spin polarization process of point defects with a high ground-state electronic spin and adjacent nuclear spins. It takes into account many microscopic features and processes that previous models have overlooked or not integrated:  the effect of the electron decoherence in the ground and excited state, the short lifetime of the excited state, the direction of the nuclear spin polarization, the anisotropy of hyperfine tensors, misaligned magnetic fields, and the simultaneous flipping of ground and excited state's spin flipping processes. We applied our model for different cases of the NV center in diamond and divacancies and related defects in 4H- and 6H-SiC, validating our model and showing its generality. Additionally, in many cases, the model provided a new insight to the microscopic mechanisms behind the examined phenomena and allowed us to draw important conclusions. We show that the electron decoherence in the ES plays an important role in DNP process, through the shortening the ES evolution time. It can limit the maximal achievable ES nuclear spin polarization, as we showed for the divacancy in SiC. By promoting fast nuclear spin rotation processes and suppressing slower ones, the electron decoherence protects the ESLAC resonance from secondary spin rotation mechanisms that appear for anisotropic hyperfine interactions. On the other hand, similar secondary processes produce more complicated fine structure at GSLAC, due to a longer evolution time permitted by the longer lifetime of the GS in the optical cycle. By studying the polarizability of remote nuclear spins, we addressed the long-standing question concerning the hyperfine parameter dependence of the maximal achievable polarization. As we showed, the strength of $A_{\perp}$ primarily governs the polarizability of remote nuclei, while the ES electron spin decoherence is an important factor as well. Throughout our investigations, electron spin decoherence appeared as an important, sometimes major, limitation of the polarizability of the nuclear spins. Therefore, any treatment that affects the point defect electron spin coherence times, like isotope enrichment, can drastically affect the DNP process as well.

\begin{acknowledgments} 

Discussions with Zolt\'an Bodrog are highly appreciated. Support from the Knut \& Alice Wallenberg Foundation ``Isotopic Control for Ultimate Materials Properties'', the Swedish Research Council (VR) Grants No.\ 621-2011-4426 and  621-2011-4249, the Swedish Foundation for Strategic Research  program SRL grant No.\ 10-0026, the Grant of Russian Federation Ministry for Science and Education (grant No. 14.Y26.31.0005), the Tomsk State University Academic D. I. Mendeleev Fund Program (project No. 8.1.18.2015), the Swedish National Infrastructure for Computing Grants No.~SNIC 2013/1-331, and the ``Lend\"ulet program" of Hungarian Academy of Sciences is acknowledged.

\end{acknowledgments}

\appendix*
\section{Spin Hamiltonian matrices}

Here, we specify the spin Hamiltonian matrices for the coupled system of a $S=1$ electron spin and an $I=1/2$ nuclear spin. The spin Hamiltonian consists of four terms (see Eq. (1) in the main text). The corresponding four matrices, as written in the basis of $\left| +1 \uparrow \right\rangle$, $\left| +1 \downarrow \right\rangle$, $\left| 0 \uparrow \right\rangle$, $\left| 0 \downarrow \right\rangle$, $\left| -1 \uparrow \right\rangle$, and $\left| -1 \downarrow \right\rangle$ states, where the spin quantization direction $z$ is parallel to the $C_{3}$ axis of the defect, are listed below. The occurrence of non-zero off-diagonal elements, which are responsible for spin rotation processes among the basis states, are discussed in each cases.

\begin{itemize}
\item The zero-field interaction term
 \begin{equation}
\hat{H}_{\text{zfs}} \! \left( D, E\right) = 
\begin{pmatrix}
 D & 0 & 0 & 0 & E & 0 \\
 0 & D & 0 & 0 & 0 & E\\
 0 & 0 & 0 & 0 & 0 & 0 \\
 0 & 0 & 0 & 0 & 0 & 0 \\
 E & 0 & 0 & 0 & D & 0 \\
 0 & E & 0 & 0 & 0 & D \\
 \end{pmatrix} \text{,}
 \end{equation}
 where $D$ and $E$ are the zero-field-splitting parameters.  For point defects of $C_{3v}$ symmetry, the $E$ parameter is zero. Thus there are no off-diagonal elements in the spin Hamiltonian due to the zero-field interaction. 
 
 \item Electron spin Zeeman term
 \begin{equation}
\hat{H}_{\text{Z}} \! \left( B, \theta_{B} \right) = g_{e} \mu_{B} B
\begin{pmatrix}
 \cos \theta_{B}   & 0 & \frac{1}{\sqrt{2}} \sin \theta_{B}  & 0 & 0 & 0 \\
 0 & \cos \theta_{B}  & 0 & \frac{1}{\sqrt{2}} \sin \theta_{B}  & 0 & 0\\
 \frac{1}{\sqrt{2}} \sin \theta_{B}   & 0 & 0 & 0 & \frac{1}{\sqrt{2}} \sin \theta_{B}   & 0 \\
 0 & \frac{1}{\sqrt{2}} \sin \theta_{B}  & 0 & 0 & 0 & \frac{1}{\sqrt{2}} \sin \theta_{B}  \\
 0 & 0 & \frac{1}{\sqrt{2}} \sin \theta_{B}  & 0 & -\cos \theta_{B}  & 0 \\
 0 & 0 & 0 & \frac{1}{\sqrt{2}} \sin \theta_{B}  & 0 & -\cos \theta_{B}  \\
 \end{pmatrix} \text{,}
 \end{equation}
where $B$ is the strength of the external magnetic field, $\theta_{B}$ is angle of the $C_{3}$ axis of the defect and the external magnetic field, $g_{e}$ is the g-factor of the electron spin, and $\mu_{\text{B}}$ is the Bohr magneton. 

As can been seen, off-diagonal matrix elements appear for $\theta_{\text{B}} \neq 0$. These elements correspond to the electron spin ladder operators $\hat{S}_{+}$ and $\hat{S}_{-}$ and couple $\left| M_S \text{,} M_I\right\rangle$ and $\left| M_S \pm 1\text{,} M_I\right\rangle$ states.

\item Nuclear spin Zeeman term
\begin{equation}
\hat{H}_{\text{NZ}} \! \left( B, \theta_{B} \right) = \frac{1}{2} g_{\text{N}} \mu_{\text{N}} B
\begin{pmatrix}
 \cos \theta_{B}  & \sin \theta_{B} & 0 & 0 & 0 & 0 \\
 \sin \theta_{B} & -\cos \theta_{B} & 0 & 0 & 0 & 0 \\
 0 & 0 & \cos \theta_{B} & \sin \theta_{B} & 0 & 0 \\
 0 & 0 & \sin \theta_{B} & -\cos \theta_{B} & 0 & 0 \\
 0 & 0 & 0 & 0 & \cos \theta_{B} & \sin \theta_{B} \\
 0 & 0 & 0 & 0 & \sin \theta_{B} & -\cos \theta_{B} \\
 \end{pmatrix} \text{,}
\end{equation}
where $B$ is the strength of the external magnetic field, $\theta_{B}$ is angle of the $C_{3}$ axis of the defect and the external magnetic field,  $g_{\text{N}}$ is the g-factor of the nuclear spin, and $\mu_{\text{N}}$ is the nuclear magneton.

Similarly to the electron spin Zeeman term, off-diagonal matrix elements are present for $\theta_{\text{B}} \neq 0$. Here, these off-diagonal elements correspond to the nuclear spin ladder operators $\hat{I}_{+}$ and $\hat{I}_{-}$ and couple $\left| M_S \text{,} M_I\right\rangle$ and $\left| M_S \text{,} M_I \pm 1 \right\rangle$ states.

\item Hyperfine interaction term
\begin{equation}
\hat{H}_{\text{hyp}} \! \left( A_{xx}, A_{yy}, A_{zz}, \theta \right) = \frac{1}{2}
\begin{pmatrix}
 a & b & \frac{1}{\sqrt{2}} b & \frac{1}{\sqrt{2}} c_{-} & 0 & 0 \\
 b & -a & \frac{1}{\sqrt{2}}  c_{+} &  -\frac{1}{\sqrt{2}} b & 0 & 0\\
 \frac{1}{\sqrt{2}} b & \frac{1}{\sqrt{2}}  c_{+}  & 0 & 0 & \frac{1}{\sqrt{2}} b & \frac{1}{\sqrt{2}} c_{-} \\
 \frac{1}{\sqrt{2}} c_{-}  & -\frac{1}{\sqrt{2}} b & 0 & 0 & \frac{1}{\sqrt{2}}  c_{+} & -\frac{1}{\sqrt{2}} b \\
 0 & 0 & \frac{1}{\sqrt{2}} b & \frac{1}{\sqrt{2}}  c_{+} & -a & -b \\
 0 & 0 & \frac{1}{\sqrt{2}} c_{-} & -\frac{1}{\sqrt{2}} b & -b & a \\
 \end{pmatrix}
 \text{,}
\end{equation}
 where
 \begin{eqnarray}
 a =  A_{zz} \cos^2 \theta  + A_{xx} \sin^2 \theta  \\
 b = \left( A_{zz} - A_{xx} \right) \cos \theta \sin \theta  \\
 c_{\pm} =  A_{xx} \cos^2 \theta  + A_{zz} \sin^2 \theta  \pm A_{yy}  \text{,}
 \end{eqnarray}
where $A_{xx}$, $A_{yy}$, and $A_{zz}$ define the hyperfine coupling strength and $\theta$ defines the polar angle of  the principal axis of the hyperfine tensor.

For the case of $\theta = 0$ and $A_{xx} = A_{yy} = A_{\perp}$, the non-zero off diagonal matrix elements $\frac{1}{2 \sqrt{2} } c_{+} = \frac{1}{\sqrt{2}} A_{\perp}$ correspond to the $\hat{S}_{\pm} \hat{I}_{\mp}$ operator combinations. These elements mix $\left| M_S \text{,} M_I\right\rangle$ and $\left| M_S \pm 1 \text{,} M_I \mp 1 \right\rangle$ states. When $A_{xx} \neq A_{yy}$ but $\theta = 0$, new off-diagonal elements are present in connection to the $\hat{S}_{\pm} \hat{I}_{\pm}$ operators. The coupling strength is $\frac{1}{2 \sqrt{2} } c_{-} = \frac{1}{2\sqrt{2}} \left( A_{xx} - A_{yy} \right)$ and the coupled states are $\left| M_S \text{,} M_I\right\rangle$ and $\left| M_S \pm 1 \text{,} M_I \pm 1 \right\rangle$. When $\theta \neq 0$ and $\theta \neq \frac{n \pi} {2}$, where $n$ is an integer, precession like spin rotation process can appear due to the $\hat{S}_{\pm} \hat{I}_{z}$ and $\hat{S}_{z} \hat{I}_{\pm}$ operators that couple state similar to the electron and nuclear spin Zeeman effects, respectively.

\end{itemize}

\bibliographystyle{apsrev4-1}

%

\end{document}